\documentstyle[twoside,fleqn,espcrc2]{article}

\newcommand{\nl}{\nonumber \\}
\newcommand{\bearray}{\begin{eqnarray}}
\newcommand{\eearray}{\end{eqnarray}}
\newcommand{\be}{\begin{equation}}
\newcommand{\ee}{\end{equation}}
\newcommand{\eq}[1]{{Eq.~(\ref{#1})}}

\newcommand{\e}{{\rm e}}
\newcommand{\half}{{\mbox{$\frac{1}{2}$}}}
\newcommand{\order}{{\cal O}}

\newcommand{\Dv}{{\bf D}}
\newcommand{\Bv}{{\bf B}}
\newcommand{\pv}{{\bf p}}
\newcommand{\si}{\sigma}
\newcommand{\Tr}{\hbox{Tr}}

\renewcommand{\Re}{ {\rm Re}\, }

\newcommand{\pl}{{\rm pl}}
\newcommand{\rt}{{\rm rt}}

\newcommand{\pg}{{\rm pg}}
\newcommand{\ls}[1]{\mbox{$\frac{1}{3}$\,Re\,Tr}(1-#1)}
\newcommand{\third}{{\textstyle {1\over 3}}}

\begin{document}

\title{QCD for Coarse Lattices\thanks{
Invited talk at {\sl Lattice '95\/}, Melbourne, Australia (1995).}}

\author{G. Peter Lepage \\
 \small Floyd R. Newman Laboratory of Nuclear Studies \\
  \small Cornell University, Ithaca, NY 14853 \\
 \small \verb+gpl@hepth.cornell.edu+}

\date{\small September 10, 1995}

\maketitle

\section{A Revolution for Lattice QCD?}

In this lecture I discuss the prospects for simulating QCD accurately on very
coarse lattices. This subject is very important because the cost of a
Monte Carlo simulation of QCD is very sensitive to the lattice
spacing. Simulation cost varies as
\be
\mbox{cost} \propto \left(\frac{L}{a}\right)^4\,
\left(\frac{1}{a^2}\right)^\delta \,
\left(\frac{1}{m_\pi^2 a}\right)^\epsilon,
\ee
where $a$ is the lattice spacing, $L$ is the length of one side of the
lattice, $m_\pi$~is the pion's mass in the simulation, and~$\delta$
and~$\epsilon$ are exponents between 0 and~1 that are related to algorithm
performance. With the best algorithms the cost of a full QCD simulation
QCD increases like $1/a^6$ as $a$ is decreased, making lattice spacing
by far the most important determinant of the cost. The very large exponent
means that we that we should do everything we can to keep~$a$ as large as
possible.

Until recently most of us felt
that lattice spacings of order .05--.1~fm or
less were necessary before accurate simulations became possible. In
this lecture I show
new Monte Carlo results from several groups indicating that accurate
results can be obtained even with lattice spacings as large as .4~fm.
Given that the cost is proportional to $1/a^6$, an $a\!=\!.4$~fm simulation
costs $10^3$--$10^6$~times less than one with conventional lattice spacings.
This is a very dramatic improvement\,---\,far larger, for example, than the
difference between a high-end PC and a typical supercomputer. If the coarse
lattices really work, lattice QCD is about to become vastly more productive.

\section{Improved Actions}
A major obstacle to using large lattice spacings is that traditional
discretizations of QCD are not very accurate when $a$~equals, say,~.4~fm.
We need improved discretizations for actions and operators. There are two
standard approaches to the design of improved actions and operators. We
examine each of them, first for classical field theory and then taking
account of quantum effects.

\subsection{Perfect Fields (Classical)}
The first approach to discretizing a classical field is based upon standard
finite-difference approximations from numerical analysis; in the context of
lattice QCD this approach is frequently associated with Symanzik's
work\,\cite{syman}.
In this approach the lattice theory is designed so that the lattice
field~$\phi_i$ at node~$x_i$ equals the corresponding continuum
field~$\phi(x)$ evaluated at that node:\footnote{The analogue of this
equation for the QCD gauge field is
\be
U_\mu(x) = {\rm P}\,\exp\,i\int_x^{x+a\hat\mu} \!\!\!\!g A\cdot dx
\nonumber \ee
where $A_\mu$ is the continuum vector potential and ${\rm
P}$~denotes path ordering.}
\be
\phi_i \, = \, \phi(x\!=\!x_i).
\ee
Given this identity the lattice version of nonlinear terms in the
field equations is exact: for example,
\be
\lambda\,\phi(x)^3 \longrightarrow \lambda\,\phi_i^3.
\ee
Continuum derivatives, however, must be approximated. Using finite
differences, for example,
\be \label{d2expansion}
\partial^2\phi(x_i) \longrightarrow \left(\Delta^{(2)} -
\frac{a^2}{12}\left(\Delta^{(2)}\right)^2 +\cdots\right) \phi_i
\ee
where
\be
\Delta^{(2)}\phi \equiv \frac{\phi_{i+1}-2\phi_i+\phi_{i-1}}{a^2}.
\ee

The expansion of $\partial^2\phi(x_i)$ is an infinite series in powers
of~$a^2$.
Normally only a couple of terms are needed. For example, if we are modelling
a bump in~$\phi(x)$ that is 3~or 4~lattice spacings across, the $a^2$~term
in this expansion  is typically only of
order 10--20\% the leading term, while the $a^4$~term is only a few
percent. The terms of order~$a^4$ and higher
would be unnecessary for most applications in lattice QCD.

When one wishes to reduce the finite-$a$ errors in a simulation,
it is usually far more efficient to improve the discretization of the
derivatives (by including more terms in the expansion) than to reduce
the lattice spacing. For example, with just the first term in the
approximation to~$\partial^2\phi$, cutting the lattice
spacing in half would reduce a 20\%~error to~5\%; but the cost
would increase by a factor of $2^6\!=\!64$ in a simulation where cost
goes like~$1/a^6$.  On
the other hand, including the $a^2$ correction to the derivative, while
working at the larger
lattice spacing, achieves the same reduction in error but at a cost
increase of only a factor of~2.

Although not  necessary in practice, it is possible to sum all  the
terms in the $a^2$~expansion of a continuum derivative using Fourier
transforms: we replace
\be
\partial^2\phi(x_i) \to -\sum_j d^{(2)}_{\rm pf}(x_i\!-\!x_j)\,\phi_j
\ee
where
\be \label{d2slac}
d^{(2)}_{\rm pf} (x_i\!-\!x_j) \equiv
\int\frac{dp}{2\pi}\,p^2\,\e^{ip(x_i-x_j)}\,\theta(|p|\!<\!\pi/a).
\ee
We do not do this because $d^{(2)}_{\rm pf}(x_i\!-\!x_j)$ falls off only as
$1/|x_i\!-\!x_j|^2$ for large~$|x_i\!-\!x_j|$, resulting in highly nonlocal
actions and operators that are very costly to simulate, particularly when
gauge fields are involved. So with ``perfect fields'' it is generally far
better to truncate the $a^2$~expansion than to sum to all orders. However
this analysis suggests a different strategy for discretization that we now
examine.

\subsection{Perfect Actions (Classical)}

The slow fall-off in the all-orders or perfect derivative~$d^{(2)}_{\rm
pf}$ of the last section is caused by the abruptness of the lattice
cutoff (the $\theta$~funtion in \eq{d2slac}).
Wilson noticed that
$d^{(2)}$ can be made to vanish faster than any power of
$1/|x_i-x_j|$ if a smoother cutoff is introduced\,\cite{wilson}.
In his approach the lattice field at node~$x_i$ is
{\em not\/} equal to the continuum field there, but rather equals the
continuum field averaged over an interval centered at that node: for
example, in one dimension\footnote{The smearing discussed here is conceptually
simple, but better smearings, resulting in improved locality, are possible.
In \cite{hasen1} the authors use  stochastic smearing in which $\phi_i$
equals the blocked continuum field only on average. This introduces a new
parameter that can be tuned to optimize the action.}
\bearray
\phi_i &=& \frac{1}{a}\,\int_{x_i-a/2}^{x_i+a/2}\phi(x)\,dx\\
&=& \int_{-\infty}^{\infty}\frac{dp}{2\pi}\,\e^{ip x_i}\,\Pi(p)\,\phi(p)
\eearray
where, in the transform, smearing function
\be
\Pi(p) \equiv \frac{2\sin(p a/2)}{p a}
\ee
provides a smooth cutoff at large~$p$.
The lattice action for the smeared fields is
designed so that tree-level Green's
functions built from these fields are {\em exactly\/}
equal to the corresponding Green's functions in the continuum theory, with
the smearing function applied at the ends: for example,
\bearray
\langle \phi_i\,\phi_j \rangle
&=& \int_{x_i-a/2}^{x_i+a/2}\!\!\!\!\!dx \int_{x_j-a/2}^{x_j+a/2}
\!\!\!\!\!dy\,\,\langle\phi(x)\,\phi(y)\rangle \\
&=& \int_{-\infty}^{\infty}\frac{dp}{2\pi}\,\e^{ip
(x_i-x_j)}\,\frac{\Pi^2(p)}{p^2+m^2}.
\eearray
By rewriting the propagator $\langle\phi_i\,\phi_j\rangle$ in the form
\be
\int_{-\pi/a}^{\pi/a}\frac{dp}{2\pi}\,\e^{ip(x_i-x_j)}\,\sum_n
\frac{ \Pi^2(p\!+\!2\pi n/a)}{(p\!+\!2\pi n/a)^2+m^2},
\ee
we find that the quadratic terms of this lattice action are
\be
S^{(2)} = \half \sum_{i,j} \phi_i\,d^{(2)}_{\rm pa}(x_i\!-\!x_j,m)\,\phi_j,
\ee
where $d^{(2)}_{\rm pa}(x_i\!-\!x_j,m)$ is
\be
\int_{-\pi/a}^{\pi/a}\!\frac{dp}{2\pi}\,\e^{ip(x_i-x_j)}\!\!\left(\!\sum_n
\frac{ \Pi^2(p\!+\!2\pi n/a)}{(p\!+\!2\pi
n/a)^2\!+\!m^2}\!\right)^{\!-1}\!\!\!\!\!\!\!.
\ee
It is easily shown that, remarkably, $d^{(2)}_{\rm pa}(x_i\!-\!x_j,m)$
vanishes faster than any power of $1/|x_i\!-\!x_j|$ as the separation
increases. So the derivative terms in such an action are both exact to all
orders in $a$ and quite local\,---\,that is, the classical action is
``perfect.'' Again, this is possible
because the lattice field~$\phi_i$ is obtained
by smearing the continuum field, which introduces a smooth UV~cutoff in
momentum space rather than the abrupt cutoff of the last Section.

The complication with this approach is that it is difficult to reconstruct
the continuum fields~$\phi(x)$ from the smeared lattice fields~$\phi_i$.
Consequently nonlinear interactions, currents, and other operators are
generally complicated functions of the~$\phi_i$, and much more difficult to
design than with the previous (``perfect field'') approach, where the
lattice field is trivially related to the continuum field. This is an
important issue in QCD where phenomenological studies involve a wide range
of currents and operators. It also means that in practice the ``perfect
action'' used for an interacting theory is not really  perfect
since the arbitrarily complex functions of~$\phi_i$ that arise in real perfect
actions must be simplified for simulations. (Although recent
numerical techniques developed by Hasenfratz, Niedermayer and their
collaborators\cite{hasen2} seem to produce accurate
approximations to the classical
perfect action even for complicated theories like~QCD.)

In general there is no clear choice between the two discretization
techniques. In the ``perfect field'' approach operators and actions are
trivial to design, but this simplicity may be at the expense of locality or
accuracy. In the ``perfect action'' approach the action has no $a^n$~errors,
but this perfection requires a very considerable (and costly) design effort,
even at the classical level. As we shall see both approaches have been very
successful  applied to QCD.

\subsection{Quantum Effects}
The techniques described in the previous two sections have long been
understood, and yet they are far from commonplace in lattice QCD analyses.
This is because of quantum effects. Quantum fluctuations with
momenta~$p\! >\!\pi/a$ are excluded or badly mutilated by the lattice in
either of the approaches we have discussed. In principle we can
compensate for this by renormalizing the couplings in our quantum opertors.
Thus, for example, $\phi\,\partial^2\phi$
is discretized by the replacement
\be \label{quantumexpansion}
\phi\,\partial^2\phi \to c_0\,\phi\Delta^{(2)}\phi - c_1
\frac{a^2}{12}\phi\left(\Delta^{(2)}\right)^2\phi +\cdots
\ee
where the coefficients~$c_0$ and~$c_1$ equal unity in a classical theory but
are renormalized in a quantum theory: $c_i = 1 + \alpha_s(\pi/a) c_i^{(1)}
+\cdots$. The renormalizations are usually context specific and
so must be recomputed for each new application. However, if $a$~is small
enough, $p\! >\! \pi/a$~physics is perturbative and the
renormalizations can be calculated perturbatively.

These considerations indicate that improved actions are practical only for
lattice spacings that satisfy the following criteria:
\begin{itemize}
\item $a\!<\!(\mbox{important length scales})$: This requirement is obvious
and applies in classical as well as quantum problems. To study the static
properties of hadrons, for example, one cannot use lattice spacings much
larger than about .4~fm, since  light hadrons have radii of order~.8~fm.
On the other hand lattice spacings much smaller than this should not be
necessary if improved operators are used.

\item perturbation theory works at distances of order~$a$ or less:
The dynamics can be perturbative at distances of order the lattice spacing
only if the coupling~$\alpha_s(\pi/a)$  is small; and  if the coupling is small
then the coefficients in an expansion like \eq{quantumexpansion} are well
approximated by their classical values. Furthermore the small quantum
corrections can be systematically computed using (analytic)
perturbation theory. However, if the lattice spacing is too large the many
extra couplings in improved actions and operators must be determined
nonperturbatively (using, for example, the Monte Carlo Renormalization
Group), making improvement very costly.
\end{itemize}

The second of these criteria has long delayed the
widespread introduction of improvement techniques into lattice QCD. Until
recently it was generally believed that perturbation theory was useless
except at very small distances ($<\!.1$~fm); and therefore improved actions
at, say, $a\!=\!.4$~fm were also useless, or, at best, prohibitively
costly to generate. This view changed
completely a few years ago when Paul Mackenzie and I noticed that
perturbation theory, done correctly,  works very well even at
distances as large as~$1/2$~fm\cite{LandM}.

Perturbation theory is indeed almost useless for lattice calculations unless
one deals with the ``tadpole problem.'' The nonlinear relation between the
link field~$U_\mu(x)$, from which lattice operators are built, and the
continuum vector potential~$A_\mu(x)$ leads to tadpole
diagrams in lattice perturbation theory that do not arise in continuum
calculations. These tadpole contributions result in large
renormalizations\,---\,often as large as a factor of two or
three\,---\,that spoil naive perturbation theory. However tadpole
contributions are generically process independent and so it is
possible to measure their contribution in one quantity and then correct for
them in all other quantities. Mackenzie and I
developed a two-part prescription for doing this\,\cite{LandM}:
\begin{itemize}
\item {\sl renormalized perturbation theory:} A
renormalized coupling constant like~$\alpha_V(q^*)$ should be used in
perturbative calculations, not the bare coupling~$\alpha_{\rm lat}$. It is
important that this coupling be {\em measured} in a simulation, for example,
by extracting its value from that of the expectation value of the plaquette;
and it is important that the scale~$q^*$ be chosen appropriately for the
quantity being studied (see
\cite{LandM} for details).\footnote{In practice, this procedure is
significantly more accurate than simply boosting the bare coupling by
dividing by the plaquette.} By measuring the coupling we automatically
include any large renormalizations of the coupling due to tadpoles.

\item {\sl tadpole improvement:} Each $U_\mu(x)$ in a lattice operator
should be replaced by
\be
\tilde{U}_\mu(x) \equiv \frac{U_\mu(x)}{u_0},
\ee
where $u_0$ is the scalar mean value of the link, defined to be the fourth root
of  the expectation value of the four-link plaquette (as measured in the
simulation). The $u_0$~cancels tadpole contributions, making lattice
operators far more continuum-like in their behavior.
\end{itemize}

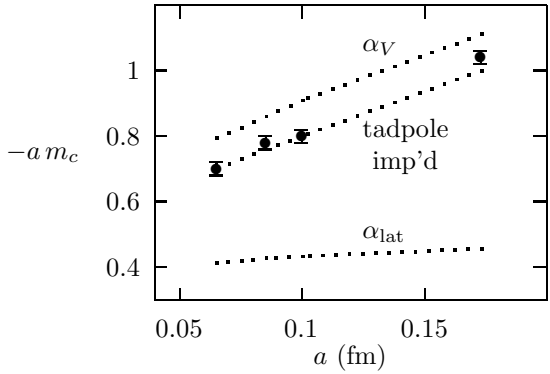
\begin{figure}
\setlength{\unitlength}{0.240900pt}
\ifx\plotpoint\undefined\newsavebox{\plotpoint}\fi
\sbox{\plotpoint}{\rule[-0.200pt]{0.400pt}{0.400pt}}%
\begin{picture}(900,600)(0,0)
\font\gnuplot=cmr10 at 10pt
\gnuplot
\sbox{\plotpoint}{\rule[-0.200pt]{0.400pt}{0.400pt}}%
\put(220.0,165.0){\rule[-0.200pt]{4.818pt}{0.400pt}}
\put(198,165){\makebox(0,0)[r]{$0.4$}}
\put(816.0,165.0){\rule[-0.200pt]{4.818pt}{0.400pt}}
\put(220.0,268.0){\rule[-0.200pt]{4.818pt}{0.400pt}}
\put(198,268){\makebox(0,0)[r]{$0.6$}}
\put(816.0,268.0){\rule[-0.200pt]{4.818pt}{0.400pt}}
\put(220.0,371.0){\rule[-0.200pt]{4.818pt}{0.400pt}}
\put(198,371){\makebox(0,0)[r]{$0.8$}}
\put(816.0,371.0){\rule[-0.200pt]{4.818pt}{0.400pt}}
\put(220.0,474.0){\rule[-0.200pt]{4.818pt}{0.400pt}}
\put(198,474){\makebox(0,0)[r]{$1$}}
\put(816.0,474.0){\rule[-0.200pt]{4.818pt}{0.400pt}}
\put(259.0,113.0){\rule[-0.200pt]{0.400pt}{4.818pt}}
\put(259,68){\makebox(0,0){$0.05$}}
\put(259.0,557.0){\rule[-0.200pt]{0.400pt}{4.818pt}}
\put(451.0,113.0){\rule[-0.200pt]{0.400pt}{4.818pt}}
\put(451,68){\makebox(0,0){$0.1$}}
\put(451.0,557.0){\rule[-0.200pt]{0.400pt}{4.818pt}}
\put(644.0,113.0){\rule[-0.200pt]{0.400pt}{4.818pt}}
\put(644,68){\makebox(0,0){$0.15$}}
\put(644.0,557.0){\rule[-0.200pt]{0.400pt}{4.818pt}}
\put(220.0,113.0){\rule[-0.200pt]{148.394pt}{0.400pt}}
\put(836.0,113.0){\rule[-0.200pt]{0.400pt}{111.778pt}}
\put(220.0,577.0){\rule[-0.200pt]{148.394pt}{0.400pt}}
\put(45,345){\makebox(0,0){$-a\,m_c$}}
\put(528,23){\makebox(0,0){$a$ (fm)}}
\put(547,216){\makebox(0,0)[l]{$\alpha_{\rm lat}$}}
\put(547,510){\makebox(0,0)[l]{$\alpha_{V}$}}
\put(547,355){\makebox(0,0)[l]{\shortstack{tadpole\\imp'd}}}
\put(220.0,113.0){\rule[-0.200pt]{0.400pt}{111.778pt}}
\put(732,495){\circle*{18}}
\put(451,371){\circle*{18}}
\put(394,360){\circle*{18}}
\put(317,319){\circle*{18}}
\put(732.0,484.0){\rule[-0.200pt]{0.400pt}{5.059pt}}
\put(722.0,484.0){\rule[-0.200pt]{4.818pt}{0.400pt}}
\put(722.0,505.0){\rule[-0.200pt]{4.818pt}{0.400pt}}
\put(451.0,360.0){\rule[-0.200pt]{0.400pt}{5.059pt}}
\put(441.0,360.0){\rule[-0.200pt]{4.818pt}{0.400pt}}
\put(441.0,381.0){\rule[-0.200pt]{4.818pt}{0.400pt}}
\put(394.0,350.0){\rule[-0.200pt]{0.400pt}{5.059pt}}
\put(384.0,350.0){\rule[-0.200pt]{4.818pt}{0.400pt}}
\put(384.0,371.0){\rule[-0.200pt]{4.818pt}{0.400pt}}
\put(317.0,309.0){\rule[-0.200pt]{0.400pt}{5.059pt}}
\put(307.0,309.0){\rule[-0.200pt]{4.818pt}{0.400pt}}
\put(307.0,330.0){\rule[-0.200pt]{4.818pt}{0.400pt}}
\sbox{\plotpoint}{\rule[-0.500pt]{1.000pt}{1.000pt}}%
\put(732,194){\usebox{\plotpoint}}
\multiput(732,194)(-20.737,-0.886){14}{\usebox{\plotpoint}}
\multiput(451,182)(-20.727,-1.091){3}{\usebox{\plotpoint}}
\multiput(394,179)(-20.647,-2.118){2}{\usebox{\plotpoint}}
\multiput(355,175)(-20.691,-1.634){2}{\usebox{\plotpoint}}
\put(317,172){\usebox{\plotpoint}}
\put(732,531){\usebox{\plotpoint}}
\multiput(732,531)(-19.465,-7.204){15}{\usebox{\plotpoint}}
\multiput(451,427)(-19.008,-8.337){3}{\usebox{\plotpoint}}
\multiput(394,402)(-18.845,-8.698){2}{\usebox{\plotpoint}}
\multiput(355,384)(-19.129,-8.054){2}{\usebox{\plotpoint}}
\put(317,368){\usebox{\plotpoint}}
\put(732,474){\usebox{\plotpoint}}
\multiput(732,474)(-19.488,-7.143){15}{\usebox{\plotpoint}}
\multiput(451,371)(-19.476,-7.175){3}{\usebox{\plotpoint}}
\multiput(394,350)(-19.254,-7.751){4}{\usebox{\plotpoint}}
\put(317,319){\usebox{\plotpoint}}
\end{picture}
\caption{The critical quark mass for Wilson's quark action plotted versus
the lattice spacing. Monte Carlo results are compared with predictions from
perturbation theory using different couplings, with and without
tadpole-improving the quark action. Results are from quenched simulations
with Wilson's gluon action and $\beta$'s~ranging from~5.7 to~6.3; we
take $a\!=\!.1$~fm at $\beta\!=\!6$.}
\label{mc}
\end{figure}

This prescription for removing tadpole contributions is now widely used,
and its success is well documented. I give here only two illustrations. In
Figure~\ref{mc} I compare (nonperturbative) Monte Carlo results for the
critical quark mass~$m_c(a)$ with results from first-order perturbation
theory\,\cite{LandM}.
The critical quark mass is the bare mass in Wilson's quark action for
which the pion is massless. This mass diverges like
$1/a$ in Wilson's theory and so should be quite perturbative. However, naive
perturbation theory, in terms of~$\alpha_{\rm lat}$, fails by almost a
factor of two even at lattice spacings as small as~$1/20$~fm. Reexpressing
perturbation theory in terms of the renormalized coupling~$\alpha_V(q^*)$
leads to excellent results, especially considering that this is
only a first-order calculation. But even these last results are significantly
improved by first tadpole-improving the quark action, and then
computing~$m_c$ using first-order renormalized perturbation theory.

\begin{figure}
\setlength{\unitlength}{0.240900pt}
\ifx\plotpoint\undefined\newsavebox{\plotpoint}\fi
\sbox{\plotpoint}{\rule[-0.200pt]{0.400pt}{0.400pt}}%
\begin{picture}(900,600)(0,0)
\font\gnuplot=cmr10 at 10pt
\gnuplot
\sbox{\plotpoint}{\rule[-0.200pt]{0.400pt}{0.400pt}}%
\put(220.0,113.0){\rule[-0.200pt]{148.394pt}{0.400pt}}
\put(220.0,113.0){\rule[-0.200pt]{0.400pt}{111.778pt}}
\put(220.0,206.0){\rule[-0.200pt]{4.818pt}{0.400pt}}
\put(198,206){\makebox(0,0)[r]{$1$}}
\put(816.0,206.0){\rule[-0.200pt]{4.818pt}{0.400pt}}
\put(220.0,299.0){\rule[-0.200pt]{4.818pt}{0.400pt}}
\put(198,299){\makebox(0,0)[r]{$2$}}
\put(816.0,299.0){\rule[-0.200pt]{4.818pt}{0.400pt}}
\put(220.0,391.0){\rule[-0.200pt]{4.818pt}{0.400pt}}
\put(198,391){\makebox(0,0)[r]{$3$}}
\put(816.0,391.0){\rule[-0.200pt]{4.818pt}{0.400pt}}
\put(220.0,484.0){\rule[-0.200pt]{4.818pt}{0.400pt}}
\put(198,484){\makebox(0,0)[r]{$4$}}
\put(816.0,484.0){\rule[-0.200pt]{4.818pt}{0.400pt}}
\put(343.0,113.0){\rule[-0.200pt]{0.400pt}{4.818pt}}
\put(343,68){\makebox(0,0){$0.1$}}
\put(343.0,557.0){\rule[-0.200pt]{0.400pt}{4.818pt}}
\put(466.0,113.0){\rule[-0.200pt]{0.400pt}{4.818pt}}
\put(466,68){\makebox(0,0){$0.2$}}
\put(466.0,557.0){\rule[-0.200pt]{0.400pt}{4.818pt}}
\put(590.0,113.0){\rule[-0.200pt]{0.400pt}{4.818pt}}
\put(590,68){\makebox(0,0){$0.3$}}
\put(590.0,557.0){\rule[-0.200pt]{0.400pt}{4.818pt}}
\put(713.0,113.0){\rule[-0.200pt]{0.400pt}{4.818pt}}
\put(713,68){\makebox(0,0){$0.4$}}
\put(713.0,557.0){\rule[-0.200pt]{0.400pt}{4.818pt}}
\put(220.0,113.0){\rule[-0.200pt]{148.394pt}{0.400pt}}
\put(836.0,113.0){\rule[-0.200pt]{0.400pt}{111.778pt}}
\put(220.0,577.0){\rule[-0.200pt]{148.394pt}{0.400pt}}
\put(45,345){\makebox(0,0){$-\ln W_{2,2}$}}
\put(528,23){\makebox(0,0){$\alpha_V(q^*)$}}
\put(220.0,113.0){\rule[-0.200pt]{0.400pt}{111.778pt}}
\put(258,145){\circle*{18}}
\put(282,165){\circle*{18}}
\put(313,188){\circle*{18}}
\put(405,255){\circle*{18}}
\put(422,267){\circle*{18}}
\put(469,301){\circle*{18}}
\put(593,374){\circle*{18}}
\put(657,445){\circle*{18}}
\put(813,511){\circle*{18}}
\put(258,145){\usebox{\plotpoint}}
\multiput(258.00,145.58)(0.600,0.496){37}{\rule{0.580pt}{0.119pt}}
\multiput(258.00,144.17)(22.796,20.000){2}{\rule{0.290pt}{0.400pt}}
\multiput(282.00,165.58)(0.675,0.496){43}{\rule{0.639pt}{0.120pt}}
\multiput(282.00,164.17)(29.673,23.000){2}{\rule{0.320pt}{0.400pt}}
\multiput(313.00,188.58)(0.697,0.499){129}{\rule{0.658pt}{0.120pt}}
\multiput(313.00,187.17)(90.635,66.000){2}{\rule{0.329pt}{0.400pt}}
\multiput(405.00,254.58)(0.712,0.492){21}{\rule{0.667pt}{0.119pt}}
\multiput(405.00,253.17)(15.616,12.000){2}{\rule{0.333pt}{0.400pt}}
\multiput(422.00,266.58)(0.785,0.497){57}{\rule{0.727pt}{0.120pt}}
\multiput(422.00,265.17)(45.492,30.000){2}{\rule{0.363pt}{0.400pt}}
\multiput(469.00,296.58)(0.796,0.499){153}{\rule{0.736pt}{0.120pt}}
\multiput(469.00,295.17)(122.473,78.000){2}{\rule{0.368pt}{0.400pt}}
\multiput(593.00,374.58)(0.802,0.498){77}{\rule{0.740pt}{0.120pt}}
\multiput(593.00,373.17)(62.464,40.000){2}{\rule{0.370pt}{0.400pt}}
\multiput(657.00,414.58)(0.797,0.499){193}{\rule{0.737pt}{0.120pt}}
\multiput(657.00,413.17)(154.471,98.000){2}{\rule{0.368pt}{0.400pt}}
\sbox{\plotpoint}{\rule[-0.500pt]{1.000pt}{1.000pt}}%
\put(258,145){\usebox{\plotpoint}}
\put(258.00,145.00){\usebox{\plotpoint}}
\put(290.25,171.12){\usebox{\plotpoint}}
\multiput(313,188)(33.556,24.437){3}{\usebox{\plotpoint}}
\multiput(405,255)(33.913,23.939){0}{\usebox{\plotpoint}}
\multiput(422,267)(34.652,22.856){2}{\usebox{\plotpoint}}
\multiput(469,298)(34.496,23.090){3}{\usebox{\plotpoint}}
\multiput(593,381)(33.708,24.227){2}{\usebox{\plotpoint}}
\multiput(657,427)(31.890,26.575){5}{\usebox{\plotpoint}}
\put(813,557){\usebox{\plotpoint}}
\put(258,145){\usebox{\plotpoint}}
\put(258.00,145.00){\usebox{\plotpoint}}
\put(290.25,171.12){\usebox{\plotpoint}}
\multiput(313,188)(33.729,24.197){3}{\usebox{\plotpoint}}
\multiput(405,254)(34.851,22.551){0}{\usebox{\plotpoint}}
\multiput(422,265)(34.991,22.334){2}{\usebox{\plotpoint}}
\multiput(469,295)(35.898,20.844){3}{\usebox{\plotpoint}}
\multiput(593,367)(36.895,19.024){2}{\usebox{\plotpoint}}
\multiput(657,400)(38.053,16.587){4}{\usebox{\plotpoint}}
\put(813,468){\usebox{\plotpoint}}
\end{picture}
\caption{Minus the logarithm of the $2\times2$~Wilson loop versus
$\alpha_V(q^*\!=\!2.65/a)$. Nonperturbative Monte Carlo results are shown
together with predictions from third-order perturbation theory. The dotted
lines indicate the possible size of fourth-order
contributions. Results are from quenched simulations
with Wilson's gluon action and $\beta$'s~ranging from~4.5 to~18.}
\label{w22}
\end{figure}
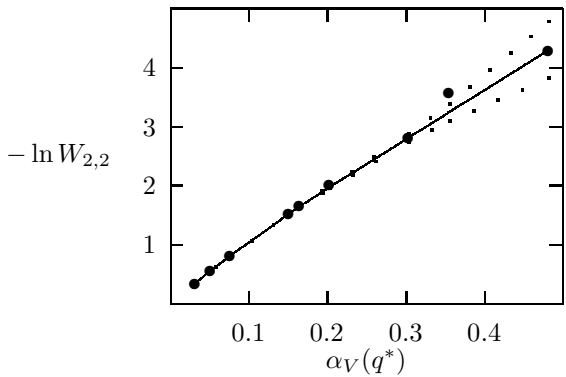

In Figure~\ref{w22} I compare Monte Carlo results for $-\ln W_{2,2}$ at
several lattice spacings with predictions from third-order (renormalized)
perturbation theory\,\cite{nrqcd-alpha}; $W_{2,2}$ is the expectation value of
a $2\times2$~Wilson loop. I plot these results versus the value of the
coupling constant~$\alpha_V(q^*\!=\!2.65/a)$ used in the perturbation theory.
This figure shows that perturbation theory is still useful even when
$\alpha_V\!\approx\!1/2$. Similar results are found for all sorts of
other quantities\,\cite{LandM}. Note that the $2\times2$~loops in these
simulations range from $10^{-6}$~fm to almost $1.2$~fm in size.

These examples and many others show that perturbation theory can usually be
trusted out to distances of order
$1/2$~fm. I say usually here because there are undoubtably situations or
quantities for which perturbation theory fails to converge or nonperturbative
effects are unusually large. However experience indicates that these are the
exception rather than the rule. Consequently our criteria for using improved
actions are satisfied if we choose lattice spacings of order $.4$~fm or less;
even classically (tree-level) improved actions, provided they are tadpole
improved, should give accurate results for such lattices.

\section{Monte Carlo Results}
In the previous sections we have argued that improved actions ought to be
very effective on coarse lattices. Here we review recent tests of this
idea. The central questions are: Do the corrections in an improved operator
really work? And, are the simulations really faster?

\subsection{Heavy Quarks\,---\,NRQCD}
One of the most thoroughly studied actions that relies upon perturbative
improvement is the NRQCD action for heavy quarks\,\cite{fiveauthors}.
In NRQCD the basic
dynamics is given by nonrelativistic Schr\"odinger theory,
\be
H_0 = -\frac{\Dv^2}{2M} + igA_0,
\ee
while relativistic effects and finite-$a$ are
included by using correction terms:
\be
\delta H = -c_1\frac{\left(\Dv^2\right)^2}{8M^3} -
c_2\frac{g\sigma\cdot\Bv}{2M} -\cdots.
\ee
The corrections were designed using the ``perfect field'' or Symanzik
approach. The couplings are computed using perturbation
theory, but only after the operators are tadpole improved; almost all work to
date has used just the tree-level values for the couplings ($c_i\!=\!1$).
Consequently there are only two parameters in the NRQCD action: the
coupling~$g$ and the bare quark mass~$M$.

\begin{figure}
\setlength{\unitlength}{.02in}
\begin{picture}(95,110)(20,-50)

\put(50,-50){\line(0,1){85}}
\multiput(48,-40)(0,20){4}{\line(1,0){4}}
\multiput(49,-40)(0,10){7}{\line(1,0){2}}
\put(47,-40){\makebox(0,0)[r]{$-40$}}
\put(47,-20){\makebox(0,0)[r]{$-20$}}
\put(47,0){\makebox(0,0)[r]{$0$}}
\put(47,20){\makebox(0,0)[r]{$20$}}
\put(47,35){\makebox(0,0)[r]{MeV}}

\multiput(100,50)(3,0){3}{\line(1,0){2}}
\put(112,50){\makebox(0,0)[l]{exp't}}
\put(105,43){\makebox(0,0)[tl]{\circle{3}}}
\put(111,43){\makebox(0,0)[l]{ $n_f = 0$}}
\put(105,36){\makebox(0,0)[tl]{\circle*{3}}}
\put(111,36){\makebox(0,0)[l]{ $n_f = 2$ }}

\put(63,-5){\makebox(0,0)[l]{$h_b$}}

\put(70,-0.8){\circle{3}}
\put(75,-2.9){\circle*{3}}
\put(75,-2.9){\line(0,1){1.2}}
\put(75,-2.9){\line(0,-1){1.2}}
\multiput(90,-40)(3,0){7}{\line(1,0){2}}
\put(110,-40){\makebox(0,0)[l]{$\chi_{b0}$}}
\put(97,-24){\circle{3}}
\put(97,-23){\line(0,1){1}}
\put(97,-25){\line(0,-1){1}}
\put(102,-34){\circle*{3}}
\put(102,-34){\line(0,1){5}}
\put(102,-34){\line(0,-1){5}}

\multiput(90,-8)(3,0){7}{\line(1,0){2}}
\put(110,-8){\makebox(0,0)[l]{$\chi_{b1}$}}
\put(97,-8.6){\circle{3}}
\put(102,-7.9){\circle*{3}}
\put(102,-7.9){\line(0,1){2.4}}
\put(102,-7.9){\line(0,-1){2.4}}

\multiput(90,13)(3,0){7}{\line(1,0){2}}
\put(110,13){\makebox(0,0)[l]{$\chi_{b2}$}}
\put(97,10.1){\circle{3}}

\put(102,11.5){\circle*{3}}
\put(102,11.5){\line(0,1){2.4}}
\put(102,11.5){\line(0,-1){2.4}}
\end{picture}
\caption{NRQCD predictions for the spin splittings between the $1P$~states
of the upsilon. Simulations used lattices with
$a^{-1}\!\approx\!2.4$~GeV.}
\label{ups-spin}
\end{figure}
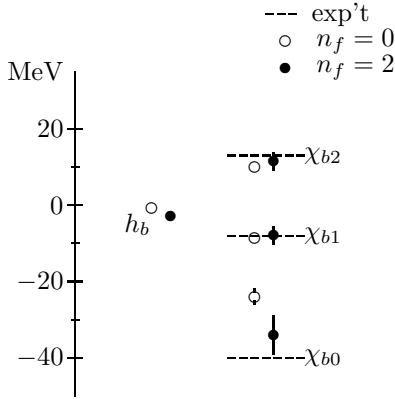

Many of the results obtained from NRQCD depend upon~$\delta H$ and so
provide nontrivial tests of the improvement program. For example,
spin-splittings in the upsilon family of mesons vanish if~$\delta H$ is
omitted. In Figure~\ref{ups-spin} I show NRQCD results for the splittings
between the $1P$~states of the upsilon\,\cite{mccallum}. The results are in
excellent agreement with experiment, particularly when extrapolated to the
correct number of light-quark flavors, $n_f\!=\!3$. Note that all of the
parameters in NRQCD are tuned from the
spin-averaged spectrum, and so there are
no free parameters in this determination
of the spin splittings. Note also that
the spin terms in~$\delta H$ all involve either chromoelectric or
chromomagnetic field operators. These operators are built from products of
four link operators and so tadpole improvement increases their magnitude by
almost a factor of two at~$\beta=6$. Without tadpole improvement the spin
splittings are much too small.

The correction terms also affect the low-momentum dispersion relation of the
upsilon,
\be
E_\Upsilon(\pv) = M_1 + \frac{\pv^2}{2\,M_2} + \cdots.
\ee
In a relativistically invariant theory $M_1\!=\!M_2$ is the upsilon mass.
However in a purely nonrelativistic theory $M_2$
equals the sum of the quark masses, which differs from $M_1$ since $M_1$
also includes the quarks' binding energy. Only when
relativistic corrections are added is $M_2$ shifted to include the
binding energy. Simulations tuned to the correct mass $M_1\!=\!9.5(1)$~GeV
give $M_2\!=\!8.2(1)$~GeV when $\delta H$  is omitted. When~$\delta H$ is
included,  the simulations give $M_2\!=\!9.5(1)$~GeV, in excellent agreement
with $M_1$\,\cite{nrqcd-upsilon}.

\subsection{Gluons\,---\,Perfect Field}
The standard Wilson action for gluons has finite-$a$
errors of order~$a^2$:
\be
\begin{array}{r@{~}c@{~}l}
1-\third \Re\Tr U_\pl &=&
  \displaystyle r_0 \sum_{\mu,\nu} \Tr(F_{\mu\nu} F_{\mu\nu}) \\[2.5ex]
  &+& a^2\Bigl[ r_1 R_1 + r_2 R_2 + r_3 R_3 \Bigr]
  \\[2.5ex]
&+& \order(a^4) + \hbox{total derivatives},
\end{array} \label{expansion}
\ee
where the $r_i$ are constants and
\be
\begin{array}{r@{~}c@{~}l}
R_1 &=& \sum_{\mu,\nu}\Tr(D_\mu F_{\mu\nu} D_\mu F_{\mu\nu}), \\[2.5ex]
R_2 &=& \sum_{\mu,\nu,\si} \Tr(D_\mu F_{\nu\si} D_\mu F_{\nu\si}), \\[2.5ex]
R_3 &=& \sum_{\mu,\nu,\si} \Tr(D_\mu F_{\mu\si} D_\nu F_{\nu\si}).
\end{array}
\ee
Only operator~$R_1$ arises at tree-level; it also breaks rotation
invariance. The other two $a^2$~operators are generated by quantum
corrections. Since operator~$R_3$ can be removed by a field redefinition,
\be \label{Atransform}
A_\mu \to A_\mu + a^2\,\alpha_s\,f(\alpha_s)\,\sum_\nu D_\nu F_{\nu\mu},
\ee
it doesn't affect spectral quantities and can be ignored.

My collaborators and I have been studying ``perfect field''
(Symanzik) techniques for improving the  gluon action; most of the
new results I
present in this section are from that collaboration and are descibed
in~\cite{large-a}. Following earlier work, we added new
Wilson loops to the gluon action
to remove~$R_1$ and~$R_2$. There are many possible
choices for the new loops, but among the simplest are the rectangle and
``parallelogram'':
\be
U_\rt =
\begin{picture}(60,30)(0,15)
  \put(10,10){\vector(0,1){12.5}}
  \put(10,10){\line(0,1){20}}
  \put(10,30){\vector(1,0){12.5}}
  \put(10,30){\vector(1,0){32.5}}
  \put(10,30){\line(1,0){40}}
  \put(50,30){\vector(0,-1){12.5}}
  \put(50,30){\line(0,-1){20}}
  \put(50,10){\vector(-1,0){12.5}}
  \put(50,10){\vector(-1,0){32.5}}
  \put(50,10){\line(-1,0){40}}
\end{picture},
\quad
U_\pg =
\begin{picture}(60,30)(0,15)
  \put(10,10){\vector(0,1){12.5}}
  \put(10,10){\line(0,1){20}}
  \put(10,30){\vector(2,1){10}}
  \put(10,30){\line(2,1){15}}
  \put(25.2,37.6){\vector(1,0){12.5}}
  \put(25.2,37.6){\line(1,0){20}}
  \put(45.2,37.6){\vector(0,-1){12.5}}
  \put(45.2,37.6){\line(0,-1){20}}
  \put(45.2,17.6){\vector(-2,-1){10}}
  \put(45.2,17.6){\line(-2,-1){15}}
  \put(30,10){\vector(-1,0){12.5}}
  \put(30,10){\line(-1,0){20}}
\end{picture}.
\ee
The improved action is\,\cite{Cur83,lw}
\bearray
S[U] &=& \beta_\pl \sum_\pl \ls{U_\pl} \nl
&+& \beta_\rt \sum_\rt \ls{U_\rt} \nl
&+& \beta_\pg \sum_\pg \ls{U_\pg} ,
\eearray
with $\beta_\pl$ given as an input, and $\beta_\rt$ and $\beta_\pg$
computed in tadpole-improved perturbation theory to cancel out~$R_1$
and~$R_2$. At tree-level, the~$\beta$'s are readily calculated by combining
expansions  like \eq{expansion} for each of the three loops.
They are tadpole-improved by dividing each
Wilson loop having $L$ links by $(u_0)^L$. One-loop
corrections have also been computed\,\cite{w}, but must be adjusted to
account for the tadpole improvement. We found:
\bearray
\label{oneloopi}
\beta_\rt &=& -\frac{\beta_\pl}{20\,u_0^2}\, \left( 1 + 0.4805\,\alpha_s
    \right), \\
\label{oneloopii}
\beta_\pg &=& -\frac{\beta_\pl}{u_0^2} \, 0.03325\,\alpha_s.
\eearray
Here $u_0$ and $\alpha_s$ are calculated from the measured
expectation value of the plaquette:
\bearray
 u_0 &=& \left(\mbox{$\frac{1}{3}$\,Re\,Tr}
   \langle U_\pl \rangle\right)^{1/4}, \\
 \alpha_s &=& -\frac{\ln\Bigl({\textstyle {1\over 3}}\mbox{Re\,Tr\,}
   \langle U_\pl\rangle\Bigr)}{3.06839}.
\eearray
As in NRQCD, there is no tuning
of the couplings for the correction terms: tadpole-improved perturbation
theory determines them in terms of the single bare
coupling~$\beta_\pl$.

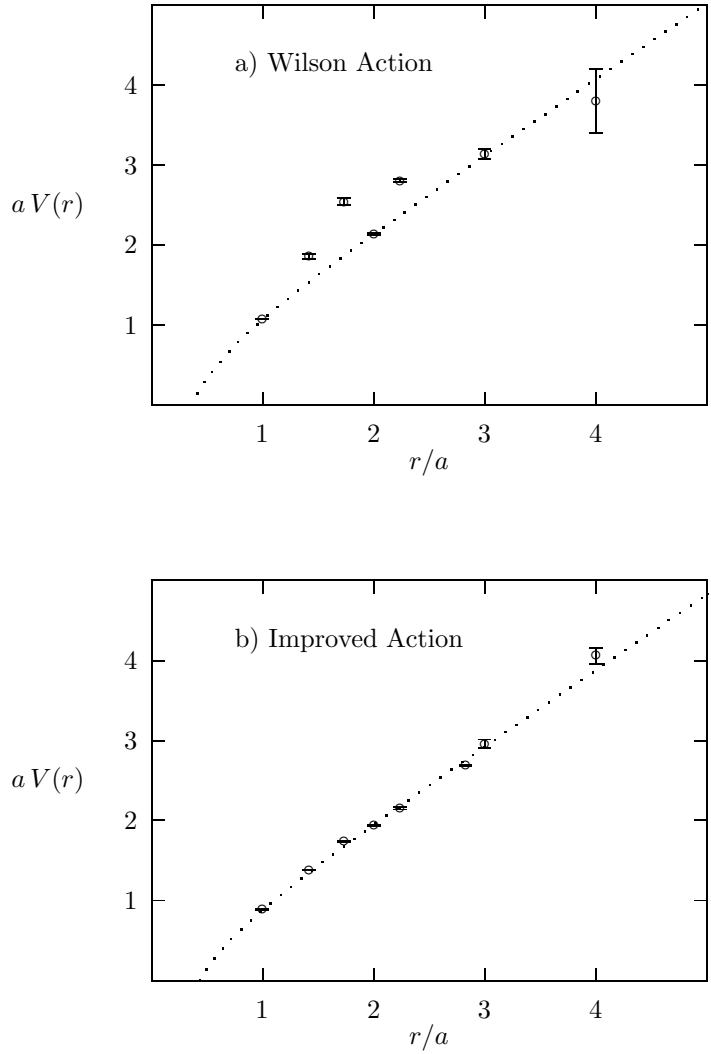
\begin{figure*}
\setlength{\unitlength}{0.240900pt}
\ifx\plotpoint\undefined\newsavebox{\plotpoint}\fi
\sbox{\plotpoint}{\rule[-0.175pt]{0.350pt}{0.350pt}}%
\begin{picture}(1200,900)(0,0)
\tenrm
\sbox{\plotpoint}{\rule[-0.175pt]{0.350pt}{0.350pt}}%
\put(264,158){\rule[-0.175pt]{210.065pt}{0.350pt}}
\put(264,158){\rule[-0.175pt]{0.350pt}{151.526pt}}
\put(264,284){\rule[-0.175pt]{4.818pt}{0.350pt}}
\put(242,284){\makebox(0,0)[r]{$1$}}
\put(1116,284){\rule[-0.175pt]{4.818pt}{0.350pt}}
\put(264,410){\rule[-0.175pt]{4.818pt}{0.350pt}}
\put(242,410){\makebox(0,0)[r]{$2$}}
\put(1116,410){\rule[-0.175pt]{4.818pt}{0.350pt}}
\put(264,535){\rule[-0.175pt]{4.818pt}{0.350pt}}
\put(242,535){\makebox(0,0)[r]{$3$}}
\put(1116,535){\rule[-0.175pt]{4.818pt}{0.350pt}}
\put(264,661){\rule[-0.175pt]{4.818pt}{0.350pt}}
\put(242,661){\makebox(0,0)[r]{$4$}}
\put(1116,661){\rule[-0.175pt]{4.818pt}{0.350pt}}
\put(438,158){\rule[-0.175pt]{0.350pt}{4.818pt}}
\put(438,113){\makebox(0,0){$1$}}
\put(438,767){\rule[-0.175pt]{0.350pt}{4.818pt}}
\put(613,158){\rule[-0.175pt]{0.350pt}{4.818pt}}
\put(613,113){\makebox(0,0){$2$}}
\put(613,767){\rule[-0.175pt]{0.350pt}{4.818pt}}
\put(787,158){\rule[-0.175pt]{0.350pt}{4.818pt}}
\put(787,113){\makebox(0,0){$3$}}
\put(787,767){\rule[-0.175pt]{0.350pt}{4.818pt}}
\put(962,158){\rule[-0.175pt]{0.350pt}{4.818pt}}
\put(962,113){\makebox(0,0){$4$}}
\put(962,767){\rule[-0.175pt]{0.350pt}{4.818pt}}
\put(264,158){\rule[-0.175pt]{210.065pt}{0.350pt}}
\put(1136,158){\rule[-0.175pt]{0.350pt}{151.526pt}}
\put(264,787){\rule[-0.175pt]{210.065pt}{0.350pt}}
\put(45,472){\makebox(0,0)[l]{\shortstack{$a\,V(r)$}}}
\put(700,68){\makebox(0,0){$r/a$}}
\put(395,693){\makebox(0,0)[l]{a) Wilson Action}}
\put(264,158){\rule[-0.175pt]{0.350pt}{151.526pt}}
\put(438,293){\circle{12}}
\put(511,392){\circle{12}}
\put(566,478){\circle{12}}
\put(613,427){\circle{12}}
\put(654,510){\circle{12}}
\put(787,553){\circle{12}}
\put(962,636){\circle{12}}
\put(438,293){\usebox{\plotpoint}}
\put(428,293){\rule[-0.175pt]{4.818pt}{0.350pt}}
\put(428,293){\rule[-0.175pt]{4.818pt}{0.350pt}}
\put(511,388){\rule[-0.175pt]{0.350pt}{1.927pt}}
\put(501,388){\rule[-0.175pt]{4.818pt}{0.350pt}}
\put(501,396){\rule[-0.175pt]{4.818pt}{0.350pt}}
\put(566,473){\rule[-0.175pt]{0.350pt}{2.409pt}}
\put(556,473){\rule[-0.175pt]{4.818pt}{0.350pt}}
\put(556,483){\rule[-0.175pt]{4.818pt}{0.350pt}}
\put(613,426){\rule[-0.175pt]{0.350pt}{0.482pt}}
\put(603,426){\rule[-0.175pt]{4.818pt}{0.350pt}}
\put(603,428){\rule[-0.175pt]{4.818pt}{0.350pt}}
\put(654,508){\rule[-0.175pt]{0.350pt}{1.204pt}}
\put(644,508){\rule[-0.175pt]{4.818pt}{0.350pt}}
\put(644,513){\rule[-0.175pt]{4.818pt}{0.350pt}}
\put(787,545){\rule[-0.175pt]{0.350pt}{3.854pt}}
\put(777,545){\rule[-0.175pt]{4.818pt}{0.350pt}}
\put(777,561){\rule[-0.175pt]{4.818pt}{0.350pt}}
\put(962,586){\rule[-0.175pt]{0.350pt}{24.090pt}}
\put(952,586){\rule[-0.175pt]{4.818pt}{0.350pt}}
\put(952,686){\rule[-0.175pt]{4.818pt}{0.350pt}}
\sbox{\plotpoint}{\rule[-0.250pt]{0.500pt}{0.500pt}}%
\put(326,158){\usebox{\plotpoint}}
\put(335,176){\usebox{\plotpoint}}
\put(346,194){\usebox{\plotpoint}}
\put(358,210){\usebox{\plotpoint}}
\put(371,226){\usebox{\plotpoint}}
\put(385,242){\usebox{\plotpoint}}
\put(399,257){\usebox{\plotpoint}}
\put(414,271){\usebox{\plotpoint}}
\put(429,285){\usebox{\plotpoint}}
\put(445,299){\usebox{\plotpoint}}
\put(461,312){\usebox{\plotpoint}}
\put(477,325){\usebox{\plotpoint}}
\put(493,338){\usebox{\plotpoint}}
\put(510,351){\usebox{\plotpoint}}
\put(526,364){\usebox{\plotpoint}}
\put(543,376){\usebox{\plotpoint}}
\put(560,388){\usebox{\plotpoint}}
\put(576,401){\usebox{\plotpoint}}
\put(593,413){\usebox{\plotpoint}}
\put(610,425){\usebox{\plotpoint}}
\put(627,437){\usebox{\plotpoint}}
\put(644,449){\usebox{\plotpoint}}
\put(660,461){\usebox{\plotpoint}}
\put(677,473){\usebox{\plotpoint}}
\put(695,485){\usebox{\plotpoint}}
\put(711,497){\usebox{\plotpoint}}
\put(728,509){\usebox{\plotpoint}}
\put(745,521){\usebox{\plotpoint}}
\put(762,533){\usebox{\plotpoint}}
\put(779,545){\usebox{\plotpoint}}
\put(796,557){\usebox{\plotpoint}}
\put(813,568){\usebox{\plotpoint}}
\put(830,580){\usebox{\plotpoint}}
\put(847,592){\usebox{\plotpoint}}
\put(864,604){\usebox{\plotpoint}}
\put(882,615){\usebox{\plotpoint}}
\put(899,627){\usebox{\plotpoint}}
\put(916,639){\usebox{\plotpoint}}
\put(932,651){\usebox{\plotpoint}}
\put(950,663){\usebox{\plotpoint}}
\put(967,674){\usebox{\plotpoint}}
\put(984,686){\usebox{\plotpoint}}
\put(1001,698){\usebox{\plotpoint}}
\put(1018,710){\usebox{\plotpoint}}
\put(1035,721){\usebox{\plotpoint}}
\put(1053,733){\usebox{\plotpoint}}
\put(1070,745){\usebox{\plotpoint}}
\put(1087,756){\usebox{\plotpoint}}
\put(1104,768){\usebox{\plotpoint}}
\put(1121,780){\usebox{\plotpoint}}
\put(1131,787){\usebox{\plotpoint}}
\end{picture}
\setlength{\unitlength}{0.240900pt}
\ifx\plotpoint\undefined\newsavebox{\plotpoint}\fi
\sbox{\plotpoint}{\rule[-0.175pt]{0.350pt}{0.350pt}}%
\begin{picture}(1200,900)(0,0)
\tenrm
\sbox{\plotpoint}{\rule[-0.175pt]{0.350pt}{0.350pt}}%
\put(264,158){\rule[-0.175pt]{210.065pt}{0.350pt}}
\put(264,158){\rule[-0.175pt]{0.350pt}{151.526pt}}
\put(264,284){\rule[-0.175pt]{4.818pt}{0.350pt}}
\put(242,284){\makebox(0,0)[r]{$1$}}
\put(1116,284){\rule[-0.175pt]{4.818pt}{0.350pt}}
\put(264,410){\rule[-0.175pt]{4.818pt}{0.350pt}}
\put(242,410){\makebox(0,0)[r]{$2$}}
\put(1116,410){\rule[-0.175pt]{4.818pt}{0.350pt}}
\put(264,535){\rule[-0.175pt]{4.818pt}{0.350pt}}
\put(242,535){\makebox(0,0)[r]{$3$}}
\put(1116,535){\rule[-0.175pt]{4.818pt}{0.350pt}}
\put(264,661){\rule[-0.175pt]{4.818pt}{0.350pt}}
\put(242,661){\makebox(0,0)[r]{$4$}}
\put(1116,661){\rule[-0.175pt]{4.818pt}{0.350pt}}
\put(438,158){\rule[-0.175pt]{0.350pt}{4.818pt}}
\put(438,113){\makebox(0,0){$1$}}
\put(438,767){\rule[-0.175pt]{0.350pt}{4.818pt}}
\put(613,158){\rule[-0.175pt]{0.350pt}{4.818pt}}
\put(613,113){\makebox(0,0){$2$}}
\put(613,767){\rule[-0.175pt]{0.350pt}{4.818pt}}
\put(787,158){\rule[-0.175pt]{0.350pt}{4.818pt}}
\put(787,113){\makebox(0,0){$3$}}
\put(787,767){\rule[-0.175pt]{0.350pt}{4.818pt}}
\put(962,158){\rule[-0.175pt]{0.350pt}{4.818pt}}
\put(962,113){\makebox(0,0){$4$}}
\put(962,767){\rule[-0.175pt]{0.350pt}{4.818pt}}
\put(264,158){\rule[-0.175pt]{210.065pt}{0.350pt}}
\put(1136,158){\rule[-0.175pt]{0.350pt}{151.526pt}}
\put(264,787){\rule[-0.175pt]{210.065pt}{0.350pt}}
\put(45,472){\makebox(0,0)[l]{\shortstack{$a\,V(r)$}}}
\put(700,68){\makebox(0,0){$r/a$}}
\put(395,693){\makebox(0,0)[l]{b) Improved Action}}
\put(264,158){\rule[-0.175pt]{0.350pt}{151.526pt}}
\put(438,270){\circle{12}}
\put(511,332){\circle{12}}
\put(566,377){\circle{12}}
\put(613,402){\circle{12}}
\put(654,429){\circle{12}}
\put(757,496){\circle{12}}
\put(787,530){\circle{12}}
\put(962,669){\circle{12}}
\put(438,269){\usebox{\plotpoint}}
\put(428,269){\rule[-0.175pt]{4.818pt}{0.350pt}}
\put(428,270){\rule[-0.175pt]{4.818pt}{0.350pt}}
\put(511,332){\usebox{\plotpoint}}
\put(501,332){\rule[-0.175pt]{4.818pt}{0.350pt}}
\put(501,332){\rule[-0.175pt]{4.818pt}{0.350pt}}
\put(566,376){\rule[-0.175pt]{0.350pt}{0.482pt}}
\put(556,376){\rule[-0.175pt]{4.818pt}{0.350pt}}
\put(556,378){\rule[-0.175pt]{4.818pt}{0.350pt}}
\put(613,401){\rule[-0.175pt]{0.350pt}{0.482pt}}
\put(603,401){\rule[-0.175pt]{4.818pt}{0.350pt}}
\put(603,403){\rule[-0.175pt]{4.818pt}{0.350pt}}
\put(654,427){\rule[-0.175pt]{0.350pt}{0.964pt}}
\put(644,427){\rule[-0.175pt]{4.818pt}{0.350pt}}
\put(644,431){\rule[-0.175pt]{4.818pt}{0.350pt}}
\put(757,495){\rule[-0.175pt]{0.350pt}{0.482pt}}
\put(747,495){\rule[-0.175pt]{4.818pt}{0.350pt}}
\put(747,497){\rule[-0.175pt]{4.818pt}{0.350pt}}
\put(787,524){\rule[-0.175pt]{0.350pt}{3.132pt}}
\put(777,524){\rule[-0.175pt]{4.818pt}{0.350pt}}
\put(777,537){\rule[-0.175pt]{4.818pt}{0.350pt}}
\put(962,656){\rule[-0.175pt]{0.350pt}{6.022pt}}
\put(952,656){\rule[-0.175pt]{4.818pt}{0.350pt}}
\put(952,681){\rule[-0.175pt]{4.818pt}{0.350pt}}
\sbox{\plotpoint}{\rule[-0.250pt]{0.500pt}{0.500pt}}%
\put(339,158){\usebox{\plotpoint}}
\put(349,175){\usebox{\plotpoint}}
\put(362,192){\usebox{\plotpoint}}
\put(375,208){\usebox{\plotpoint}}
\put(388,223){\usebox{\plotpoint}}
\put(404,238){\usebox{\plotpoint}}
\put(419,252){\usebox{\plotpoint}}
\put(434,266){\usebox{\plotpoint}}
\put(450,279){\usebox{\plotpoint}}
\put(466,292){\usebox{\plotpoint}}
\put(482,305){\usebox{\plotpoint}}
\put(498,318){\usebox{\plotpoint}}
\put(515,331){\usebox{\plotpoint}}
\put(531,344){\usebox{\plotpoint}}
\put(548,356){\usebox{\plotpoint}}
\put(565,368){\usebox{\plotpoint}}
\put(582,380){\usebox{\plotpoint}}
\put(598,392){\usebox{\plotpoint}}
\put(615,405){\usebox{\plotpoint}}
\put(632,417){\usebox{\plotpoint}}
\put(649,429){\usebox{\plotpoint}}
\put(666,441){\usebox{\plotpoint}}
\put(683,453){\usebox{\plotpoint}}
\put(700,465){\usebox{\plotpoint}}
\put(717,477){\usebox{\plotpoint}}
\put(734,489){\usebox{\plotpoint}}
\put(751,501){\usebox{\plotpoint}}
\put(768,512){\usebox{\plotpoint}}
\put(785,524){\usebox{\plotpoint}}
\put(802,536){\usebox{\plotpoint}}
\put(819,548){\usebox{\plotpoint}}
\put(836,559){\usebox{\plotpoint}}
\put(853,571){\usebox{\plotpoint}}
\put(870,583){\usebox{\plotpoint}}
\put(887,595){\usebox{\plotpoint}}
\put(904,607){\usebox{\plotpoint}}
\put(921,618){\usebox{\plotpoint}}
\put(938,630){\usebox{\plotpoint}}
\put(956,642){\usebox{\plotpoint}}
\put(973,654){\usebox{\plotpoint}}
\put(990,665){\usebox{\plotpoint}}
\put(1007,677){\usebox{\plotpoint}}
\put(1024,689){\usebox{\plotpoint}}
\put(1041,700){\usebox{\plotpoint}}
\put(1059,712){\usebox{\plotpoint}}
\put(1075,724){\usebox{\plotpoint}}
\put(1093,735){\usebox{\plotpoint}}
\put(1110,747){\usebox{\plotpoint}}
\put(1127,759){\usebox{\plotpoint}}
\put(1136,765){\usebox{\plotpoint}}
\end{picture}
\caption{Static-quark potential computed on $6^4$ lattices with
$a\!\approx\!0.4$~fm
using the $\beta=4.5$ Wilson action and the
improved action with $\beta_\pl = 6.8$.}
\label{potl}
\end{figure*}

The effects of improvement are immediately apparent if one computes the
static quark potential using a lattice spacing of~.4~fm.
Our results for both the Wilson theory and the improved theory are shown in
Figure~\ref{potl}. Off-axis points deviate by as much as~40\%
{} from a fit to the on-axis points for the Wilson theory; these errors are
reduced to 4--5\% in the improved theory. The potential from the improved
theory is essentially unchanged if the $\alpha_s$ corrections are dropped
from the action, leaving just the tree-level improved theory (with
tadpole-improved operators). This suggests that
$\order(\alpha_s a^2)$~corrections, together with
$\order(a^4)$~corrections, are relatively unimportant. Our
simulations also show that the errors are 2--3~times larger without tadpole
improvement even if $\alpha_s$ corrections are retained.

Morningstar and Peardon, in their presentation to this conference, show the
first glueball calculations with this improved action. They find a dramatic
reduction in the lattice-spacing dependence of the glueball mass relative to
the Wilson theory. Also Bliss, Hornbostel and I have completed preliminary
measurements of the temperature~$T_c$ of the deconfining phase transition in
gluonic QCD\,\cite{bliss}. Results obtained from lattices with $a$
equal to
$1/2T_c$ and $1/3T_c$ agree to within the statistical errors of a few percent,
showing no sign of finite-$a$ errors (see Table~\ref{tc-cornell}).
\begin{table}
\begin{tabular}{ccc} \hline
$N_t$ & $\beta_\pl^c$ & $T_c$ (MeV) \\ \hline
2 & 6.96 & 273 (10) \\
3 & 7.45 & 285 (15) \\ \hline
\end{tabular}
\caption{The critical temperature for the QCD deconfining phase transition
computed on lattices with $a\,T_c=1/N_t$ using the improved gluon action.
The scale is set using the $1P\!-\!1S$~splitting in charmonium.}
\label{tc-cornell}
\end{table}

The spin-averaged spectrum of the $\psi$~family of
mesons is also  independent of the lattice spacing for spacings
of~.4~fm and less\,\cite{large-a}. This is illustrated by the simulation
results for the spectrum in Figure~\ref{psi-spect}; the scale for each result
is set by the $1P\!-\!1S$~splitting. The $a$-independence (scaling) of these
results is also evident if they are compared with the slope of the static
potential~$V^\prime$ evaluated at~.6~fm, as shown in
Figure~\ref{psi-scaling}. The results shown for the Wilson theory, however,
change by 40\% as $a$ is decreased, showing large $a^2$ errors.
\begin{figure}
\setlength{\unitlength}{.02in}
\begin{picture}(100,220)(-10,280)
\put(89,340){\circle{3}}\put(95,340){\makebox(0,0)[l]{$0.40$~fm}}
\put(88,329){\framebox(2,2){\mbox{}}}
             \put(95,330){\makebox(0,0)[l]{$0.33$~fm}}
\put(89,320){{\circle*{3}}}\put(95,320){\makebox(0,0)[l]{$0.24$~fm}}
\put(88,310){\rule[-\unitlength]{2\unitlength}{2\unitlength}}
         \put(95,310){\makebox(0,0)[l]{$0.17$~fm}}

\put(15,290){\line(0,1){120}}
\multiput(13,300)(0,50){3}{\line(1,0){4}}
\multiput(14,310)(0,10){9}{\line(1,0){2}}
\put(12,300){\makebox(0,0)[r]{3.0}}
\put(12,350){\makebox(0,0)[r]{3.5}}
\put(12,400){\makebox(0,0)[r]{4.0}}
\put(12,410){\makebox(0,0)[r]{GeV}}

\put(30,290){\makebox(0,0)[t]{$S$}}

\multiput(23,307)(3,0){6}{\line(1,0){2}}
\put(26,307){\circle{3}}
\put(29,306){\framebox(2,2){\mbox{}}}
\put(34,307){\circle*{3}}
\put(37,306){\rule{2\unitlength}{2\unitlength}}

\multiput(23,366)(3,0){6}{\line(1,0){2}}
\put(26,371){\circle{3}}
\put(26,369){\line(0,1){4}}
\put(29,371){\framebox(2,2){\mbox{}}}
\put(30,370){\line(0,1){4}}
\put(34,372){\circle*{3}}
\put(34,368){\line(0,1){8}}
\put(37,369){\rule{2\unitlength}{2\unitlength}}
\put(38,361){\line(0,1){16}}

\put(50,290){\makebox(0,0)[t]{$P$}}

\multiput(43,352)(3,0){6}{\line(1,0){2}}
\put(46,352){\circle{3}}
\put(49,351){\framebox(2,2){}}
\put(54,352){\circle*{3}}
\put(54,351){\line(0,1){2}}
\put(57,351){\rule{2\unitlength}{2\unitlength}}
\put(58,351.5){\line(0,1){1}}

\put(70,290){\makebox(0,0)[t]{$D$}}

\put(66,387){\circle{3}}
\put(66,382){\line(0,1){10}}
\put(69,382){\framebox(2,2){}}
\put(70,376){\line(0,1){12}}
\put(74,387){\circle*{3}}
\put(74,384){\line(0,1){6}}
\put(77,382){\rule{2\unitlength}{2\unitlength}}
\put(78,378){\line(0,1){10}}
\end{picture}


\caption{$S$, $P$, and $D$ states of charmonium computed on lattices with:
$a=0.40$~fm (improved action, $\beta_\pl=6.8$);
$a=0.33$~fm (improved action, $\beta_\pl=7.1$);
$a=0.24$~fm (improved action, $\beta_\pl=7.4$); and
$a=0.17$~fm (Wilson action, $\beta=5.7$, from NRQCD Collaboration). The
dashed lines indicate the true masses.}
\label{psi-spect}
\end{figure}
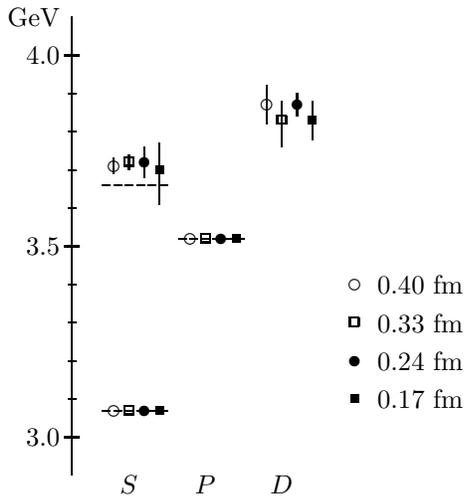
\begin{figure}
\setlength{\unitlength}{0.240900pt}
\ifx\plotpoint\undefined\newsavebox{\plotpoint}\fi
\sbox{\plotpoint}{\rule[-0.200pt]{0.400pt}{0.400pt}}%
\begin{picture}(900,600)(0,0)
\font\gnuplot=cmr10 at 10pt
\gnuplot
\sbox{\plotpoint}{\rule[-0.200pt]{0.400pt}{0.400pt}}%
\put(220.0,113.0){\rule[-0.200pt]{148.394pt}{0.400pt}}
\put(220.0,113.0){\rule[-0.200pt]{0.400pt}{111.778pt}}
\put(220.0,229.0){\rule[-0.200pt]{4.818pt}{0.400pt}}
\put(198,229){\makebox(0,0)[r]{$0.5$}}
\put(816.0,229.0){\rule[-0.200pt]{4.818pt}{0.400pt}}
\put(220.0,345.0){\rule[-0.200pt]{4.818pt}{0.400pt}}
\put(198,345){\makebox(0,0)[r]{$1$}}
\put(816.0,345.0){\rule[-0.200pt]{4.818pt}{0.400pt}}
\put(220.0,461.0){\rule[-0.200pt]{4.818pt}{0.400pt}}
\put(198,461){\makebox(0,0)[r]{$1.5$}}
\put(816.0,461.0){\rule[-0.200pt]{4.818pt}{0.400pt}}
\put(343.0,113.0){\rule[-0.200pt]{0.400pt}{4.818pt}}
\put(343,68){\makebox(0,0){$0.1$}}
\put(343.0,557.0){\rule[-0.200pt]{0.400pt}{4.818pt}}
\put(466.0,113.0){\rule[-0.200pt]{0.400pt}{4.818pt}}
\put(466,68){\makebox(0,0){$0.2$}}
\put(466.0,557.0){\rule[-0.200pt]{0.400pt}{4.818pt}}
\put(590.0,113.0){\rule[-0.200pt]{0.400pt}{4.818pt}}
\put(590,68){\makebox(0,0){$0.3$}}
\put(590.0,557.0){\rule[-0.200pt]{0.400pt}{4.818pt}}
\put(713.0,113.0){\rule[-0.200pt]{0.400pt}{4.818pt}}
\put(713,68){\makebox(0,0){$0.4$}}
\put(713.0,557.0){\rule[-0.200pt]{0.400pt}{4.818pt}}
\put(220.0,113.0){\rule[-0.200pt]{148.394pt}{0.400pt}}
\put(836.0,113.0){\rule[-0.200pt]{0.400pt}{111.778pt}}
\put(220.0,577.0){\rule[-0.200pt]{148.394pt}{0.400pt}}
\put(45,345){\makebox(0,0){$\frac{(1P\!-\!1S)}{\sqrt{V^\prime}}$}}
\put(528,23){\makebox(0,0){$a$ (fm)}}
\put(528,403){\makebox(0,0)[l]{Wilson}}
\put(528,275){\makebox(0,0)[l]{Improved}}
\put(220.0,113.0){\rule[-0.200pt]{0.400pt}{111.778pt}}
\put(429,315){\circle*{18}}
\put(725,433){\circle*{18}}
\put(429.0,308.0){\rule[-0.200pt]{0.400pt}{3.373pt}}
\put(419.0,308.0){\rule[-0.200pt]{4.818pt}{0.400pt}}
\put(419.0,322.0){\rule[-0.200pt]{4.818pt}{0.400pt}}
\put(725.0,422.0){\rule[-0.200pt]{0.400pt}{5.541pt}}
\put(715.0,422.0){\rule[-0.200pt]{4.818pt}{0.400pt}}
\put(715.0,445.0){\rule[-0.200pt]{4.818pt}{0.400pt}}
\put(516,319){\circle{18}}
\put(627,326){\circle{18}}
\put(713,322){\circle{18}}
\end{picture}
\caption{$1P\!-\!1S$~charmonium splitting divided by the square root of the
slope of the static-quark potential evaluated at~.6~fm. Results are plotted
versus lattice spacing for both the Wilson and improved gluon actions.}
\label{psi-scaling}
\end{figure}
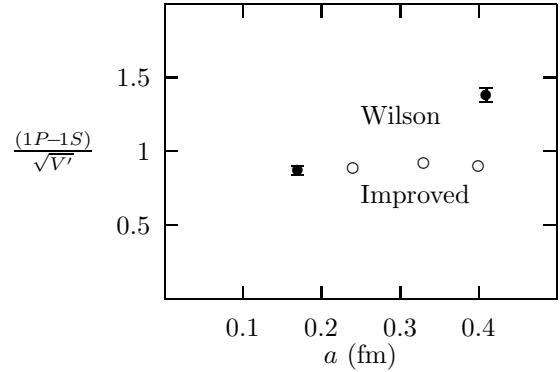

\begin{figure*}
\setlength{\unitlength}{0.240900pt}
\ifx\plotpoint\undefined\newsavebox{\plotpoint}\fi
\sbox{\plotpoint}{\rule[-0.175pt]{0.350pt}{0.350pt}}%
\begin{picture}(1200,900)(0,0)
\tenrm
\sbox{\plotpoint}{\rule[-0.175pt]{0.350pt}{0.350pt}}%
\put(264,158){\rule[-0.175pt]{210.065pt}{0.350pt}}
\put(264,158){\rule[-0.175pt]{0.350pt}{151.526pt}}
\put(264,420){\rule[-0.175pt]{4.818pt}{0.350pt}}
\put(242,420){\makebox(0,0)[r]{$5$}}
\put(1116,420){\rule[-0.175pt]{4.818pt}{0.350pt}}
\put(264,682){\rule[-0.175pt]{4.818pt}{0.350pt}}
\put(242,682){\makebox(0,0)[r]{$10$}}
\put(1116,682){\rule[-0.175pt]{4.818pt}{0.350pt}}
\put(264,158){\rule[-0.175pt]{0.350pt}{4.818pt}}
\put(264,113){\makebox(0,0){$0$}}
\put(264,767){\rule[-0.175pt]{0.350pt}{4.818pt}}
\put(700,158){\rule[-0.175pt]{0.350pt}{4.818pt}}
\put(700,113){\makebox(0,0){$0.5$}}
\put(700,767){\rule[-0.175pt]{0.350pt}{4.818pt}}
\put(1136,158){\rule[-0.175pt]{0.350pt}{4.818pt}}
\put(1136,113){\makebox(0,0){$1$}}
\put(1136,767){\rule[-0.175pt]{0.350pt}{4.818pt}}
\put(264,158){\rule[-0.175pt]{210.065pt}{0.350pt}}
\put(1136,158){\rule[-0.175pt]{0.350pt}{151.526pt}}
\put(264,787){\rule[-0.175pt]{210.065pt}{0.350pt}}
\put(45,472){\makebox(0,0)[l]{\shortstack{$R_{1S}(r)$}}}
\put(700,68){\makebox(0,0){$r$ (fm)}}
\put(700,832){\makebox(0,0){$1S$ Radial Wavefunction}}
\put(264,158){\rule[-0.175pt]{0.350pt}{151.526pt}}
\put(1006,722){\makebox(0,0)[r]{$a=0.40$~fm}}
\put(1050,722){\circle*{18}}
\put(867,200){\circle*{18}}
\put(1116,167){\circle*{18}}
\put(264,619){\circle*{18}}
\put(612,348){\circle*{18}}
\put(960,189){\circle*{18}}
\put(756,243){\circle*{18}}
\put(1042,175){\circle*{18}}
\put(1006,677){\makebox(0,0)[r]{$a=0.24$~fm}}
\put(1050,677){\circle{18}}
\put(861,199){\circle{18}}
\put(1025,172){\circle{18}}
\put(630,319){\circle{18}}
\put(781,226){\circle{18}}
\put(964,179){\circle{18}}
\put(897,190){\circle{18}}
\put(1054,169){\circle{18}}
\put(264,731){\circle{18}}
\put(475,512){\circle{18}}
\put(686,284){\circle{18}}
\put(897,192){\circle{18}}
\put(1108,172){\circle{18}}
\put(562,390){\circle{18}}
\put(736,249){\circle{18}}
\put(931,184){\circle{18}}
\put(1134,169){\circle{18}}
\put(995,174){\circle{18}}
\put(1134,164){\circle{18}}
\sbox{\plotpoint}{\rule[-0.250pt]{0.500pt}{0.500pt}}%
\put(1006,632){\makebox(0,0)[r]{quark model}}
\put(1028,632){\usebox{\plotpoint}}
\put(1048,632){\usebox{\plotpoint}}
\put(1069,632){\usebox{\plotpoint}}
\put(1090,632){\usebox{\plotpoint}}
\put(1094,632){\usebox{\plotpoint}}
\put(264,738){\usebox{\plotpoint}}
\put(264,738){\usebox{\plotpoint}}
\put(279,724){\usebox{\plotpoint}}
\put(295,711){\usebox{\plotpoint}}
\put(309,696){\usebox{\plotpoint}}
\put(324,680){\usebox{\plotpoint}}
\put(338,665){\usebox{\plotpoint}}
\put(352,650){\usebox{\plotpoint}}
\put(365,634){\usebox{\plotpoint}}
\put(378,617){\usebox{\plotpoint}}
\put(391,601){\usebox{\plotpoint}}
\put(404,585){\usebox{\plotpoint}}
\put(417,569){\usebox{\plotpoint}}
\put(429,553){\usebox{\plotpoint}}
\put(442,536){\usebox{\plotpoint}}
\put(455,520){\usebox{\plotpoint}}
\put(468,504){\usebox{\plotpoint}}
\put(481,487){\usebox{\plotpoint}}
\put(494,471){\usebox{\plotpoint}}
\put(507,455){\usebox{\plotpoint}}
\put(521,440){\usebox{\plotpoint}}
\put(534,424){\usebox{\plotpoint}}
\put(549,409){\usebox{\plotpoint}}
\put(563,394){\usebox{\plotpoint}}
\put(577,379){\usebox{\plotpoint}}
\put(592,364){\usebox{\plotpoint}}
\put(607,350){\usebox{\plotpoint}}
\put(622,336){\usebox{\plotpoint}}
\put(638,323){\usebox{\plotpoint}}
\put(654,309){\usebox{\plotpoint}}
\put(671,297){\usebox{\plotpoint}}
\put(688,285){\usebox{\plotpoint}}
\put(705,273){\usebox{\plotpoint}}
\put(722,262){\usebox{\plotpoint}}
\put(740,252){\usebox{\plotpoint}}
\put(759,242){\usebox{\plotpoint}}
\put(777,233){\usebox{\plotpoint}}
\put(797,225){\usebox{\plotpoint}}
\put(816,217){\usebox{\plotpoint}}
\put(835,210){\usebox{\plotpoint}}
\put(855,204){\usebox{\plotpoint}}
\put(875,198){\usebox{\plotpoint}}
\put(895,193){\usebox{\plotpoint}}
\put(915,188){\usebox{\plotpoint}}
\put(936,184){\usebox{\plotpoint}}
\put(956,180){\usebox{\plotpoint}}
\put(977,177){\usebox{\plotpoint}}
\put(997,174){\usebox{\plotpoint}}
\put(1018,172){\usebox{\plotpoint}}
\put(1038,170){\usebox{\plotpoint}}
\put(1059,168){\usebox{\plotpoint}}
\put(1080,166){\usebox{\plotpoint}}
\put(1101,165){\usebox{\plotpoint}}
\put(1121,164){\usebox{\plotpoint}}
\put(1136,164){\usebox{\plotpoint}}
\end{picture}
\setlength{\unitlength}{0.240900pt}
\ifx\plotpoint\undefined\newsavebox{\plotpoint}\fi
\sbox{\plotpoint}{\rule[-0.175pt]{0.350pt}{0.350pt}}%
\begin{picture}(1200,900)(0,0)
\tenrm
\sbox{\plotpoint}{\rule[-0.175pt]{0.350pt}{0.350pt}}%
\put(264,158){\rule[-0.175pt]{210.065pt}{0.350pt}}
\put(264,158){\rule[-0.175pt]{0.350pt}{151.526pt}}
\put(264,368){\rule[-0.175pt]{4.818pt}{0.350pt}}
\put(242,368){\makebox(0,0)[r]{$2$}}
\put(1116,368){\rule[-0.175pt]{4.818pt}{0.350pt}}
\put(264,577){\rule[-0.175pt]{4.818pt}{0.350pt}}
\put(242,577){\makebox(0,0)[r]{$4$}}
\put(1116,577){\rule[-0.175pt]{4.818pt}{0.350pt}}
\put(264,787){\rule[-0.175pt]{4.818pt}{0.350pt}}
\put(242,787){\makebox(0,0)[r]{$6$}}
\put(1116,787){\rule[-0.175pt]{4.818pt}{0.350pt}}
\put(264,158){\rule[-0.175pt]{0.350pt}{4.818pt}}
\put(264,113){\makebox(0,0){$0$}}
\put(264,767){\rule[-0.175pt]{0.350pt}{4.818pt}}
\put(700,158){\rule[-0.175pt]{0.350pt}{4.818pt}}
\put(700,113){\makebox(0,0){$0.5$}}
\put(700,767){\rule[-0.175pt]{0.350pt}{4.818pt}}
\put(1136,158){\rule[-0.175pt]{0.350pt}{4.818pt}}
\put(1136,113){\makebox(0,0){$1$}}
\put(1136,767){\rule[-0.175pt]{0.350pt}{4.818pt}}
\put(264,158){\rule[-0.175pt]{210.065pt}{0.350pt}}
\put(1136,158){\rule[-0.175pt]{0.350pt}{151.526pt}}
\put(264,787){\rule[-0.175pt]{210.065pt}{0.350pt}}
\put(45,472){\makebox(0,0)[l]{\shortstack{$R_{1P}(r)$}}}
\put(700,68){\makebox(0,0){$r$ (fm)}}
\put(700,832){\makebox(0,0){$1P$ Radial Wavefunction}}
\put(264,158){\rule[-0.175pt]{0.350pt}{151.526pt}}
\put(1006,722){\makebox(0,0)[r]{$a=0.40$~fm}}
\put(1050,722){\circle*{18}}
\put(612,521){\circle*{18}}
\put(756,411){\circle*{18}}
\put(1042,243){\circle*{18}}
\put(867,321){\circle*{18}}
\put(1116,214){\circle*{18}}
\put(960,299){\circle*{18}}
\put(1042,244){\circle*{18}}
\put(1116,211){\circle*{18}}
\put(1028,722){\rule[-0.175pt]{15.899pt}{0.350pt}}
\put(1028,712){\rule[-0.175pt]{0.350pt}{4.818pt}}
\put(1094,712){\rule[-0.175pt]{0.350pt}{4.818pt}}
\put(612,503){\rule[-0.175pt]{0.350pt}{8.431pt}}
\put(602,503){\rule[-0.175pt]{4.818pt}{0.350pt}}
\put(602,538){\rule[-0.175pt]{4.818pt}{0.350pt}}
\put(756,404){\rule[-0.175pt]{0.350pt}{3.613pt}}
\put(746,404){\rule[-0.175pt]{4.818pt}{0.350pt}}
\put(746,419){\rule[-0.175pt]{4.818pt}{0.350pt}}
\put(1042,238){\rule[-0.175pt]{0.350pt}{2.650pt}}
\put(1032,238){\rule[-0.175pt]{4.818pt}{0.350pt}}
\put(1032,249){\rule[-0.175pt]{4.818pt}{0.350pt}}
\put(867,314){\rule[-0.175pt]{0.350pt}{3.373pt}}
\put(857,314){\rule[-0.175pt]{4.818pt}{0.350pt}}
\put(857,328){\rule[-0.175pt]{4.818pt}{0.350pt}}
\put(1116,211){\rule[-0.175pt]{0.350pt}{1.445pt}}
\put(1106,211){\rule[-0.175pt]{4.818pt}{0.350pt}}
\put(1106,217){\rule[-0.175pt]{4.818pt}{0.350pt}}
\put(960,289){\rule[-0.175pt]{0.350pt}{4.577pt}}
\put(950,289){\rule[-0.175pt]{4.818pt}{0.350pt}}
\put(950,308){\rule[-0.175pt]{4.818pt}{0.350pt}}
\put(1042,240){\rule[-0.175pt]{0.350pt}{1.927pt}}
\put(1032,240){\rule[-0.175pt]{4.818pt}{0.350pt}}
\put(1032,248){\rule[-0.175pt]{4.818pt}{0.350pt}}
\put(1116,208){\rule[-0.175pt]{0.350pt}{1.686pt}}
\put(1106,208){\rule[-0.175pt]{4.818pt}{0.350pt}}
\put(1106,215){\rule[-0.175pt]{4.818pt}{0.350pt}}
\put(1006,677){\makebox(0,0)[r]{$a=0.24$~fm}}
\put(1050,677){\circle{18}}
\put(1054,231){\circle{18}}
\put(897,286){\circle{18}}
\put(931,265){\circle{18}}
\put(1025,222){\circle{18}}
\put(995,251){\circle{18}}
\put(1134,205){\circle{18}}
\put(475,555){\circle{18}}
\put(562,560){\circle{18}}
\put(736,442){\circle{18}}
\put(931,295){\circle{18}}
\put(964,248){\circle{18}}
\put(1134,246){\circle{18}}
\put(1054,214){\circle{18}}
\put(630,521){\circle{18}}
\put(781,399){\circle{18}}
\put(1108,164){\circle{18}}
\put(964,275){\circle{18}}
\put(1134,162){\circle{18}}
\put(1134,194){\circle{18}}
\put(686,480){\circle{18}}
\put(736,430){\circle{18}}
\put(861,329){\circle{18}}
\put(897,312){\circle{18}}
\put(1025,242){\circle{18}}
\put(1054,238){\circle{18}}
\put(781,389){\circle{18}}
\put(897,304){\circle{18}}
\put(1028,677){\rule[-0.175pt]{15.899pt}{0.350pt}}
\put(1028,667){\rule[-0.175pt]{0.350pt}{4.818pt}}
\put(1094,667){\rule[-0.175pt]{0.350pt}{4.818pt}}
\put(1054,221){\rule[-0.175pt]{0.350pt}{4.577pt}}
\put(1044,221){\rule[-0.175pt]{4.818pt}{0.350pt}}
\put(1044,240){\rule[-0.175pt]{4.818pt}{0.350pt}}
\put(897,264){\rule[-0.175pt]{0.350pt}{10.840pt}}
\put(887,264){\rule[-0.175pt]{4.818pt}{0.350pt}}
\put(887,309){\rule[-0.175pt]{4.818pt}{0.350pt}}
\put(931,254){\rule[-0.175pt]{0.350pt}{5.300pt}}
\put(921,254){\rule[-0.175pt]{4.818pt}{0.350pt}}
\put(921,276){\rule[-0.175pt]{4.818pt}{0.350pt}}
\put(1025,212){\rule[-0.175pt]{0.350pt}{4.818pt}}
\put(1015,212){\rule[-0.175pt]{4.818pt}{0.350pt}}
\put(1015,232){\rule[-0.175pt]{4.818pt}{0.350pt}}
\put(995,236){\rule[-0.175pt]{0.350pt}{6.986pt}}
\put(985,236){\rule[-0.175pt]{4.818pt}{0.350pt}}
\put(985,265){\rule[-0.175pt]{4.818pt}{0.350pt}}
\put(1134,197){\rule[-0.175pt]{0.350pt}{3.854pt}}
\put(1124,197){\rule[-0.175pt]{4.818pt}{0.350pt}}
\put(1124,213){\rule[-0.175pt]{4.818pt}{0.350pt}}
\put(475,483){\rule[-0.175pt]{0.350pt}{34.690pt}}
\put(465,483){\rule[-0.175pt]{4.818pt}{0.350pt}}
\put(465,627){\rule[-0.175pt]{4.818pt}{0.350pt}}
\put(562,522){\rule[-0.175pt]{0.350pt}{18.308pt}}
\put(552,522){\rule[-0.175pt]{4.818pt}{0.350pt}}
\put(552,598){\rule[-0.175pt]{4.818pt}{0.350pt}}
\put(736,407){\rule[-0.175pt]{0.350pt}{17.104pt}}
\put(726,407){\rule[-0.175pt]{4.818pt}{0.350pt}}
\put(726,478){\rule[-0.175pt]{4.818pt}{0.350pt}}
\put(931,265){\rule[-0.175pt]{0.350pt}{14.213pt}}
\put(921,265){\rule[-0.175pt]{4.818pt}{0.350pt}}
\put(921,324){\rule[-0.175pt]{4.818pt}{0.350pt}}
\put(964,237){\rule[-0.175pt]{0.350pt}{5.300pt}}
\put(954,237){\rule[-0.175pt]{4.818pt}{0.350pt}}
\put(954,259){\rule[-0.175pt]{4.818pt}{0.350pt}}
\put(1134,207){\rule[-0.175pt]{0.350pt}{18.790pt}}
\put(1124,207){\rule[-0.175pt]{4.818pt}{0.350pt}}
\put(1124,285){\rule[-0.175pt]{4.818pt}{0.350pt}}
\put(1054,207){\rule[-0.175pt]{0.350pt}{3.373pt}}
\put(1044,207){\rule[-0.175pt]{4.818pt}{0.350pt}}
\put(1044,221){\rule[-0.175pt]{4.818pt}{0.350pt}}
\put(630,483){\rule[-0.175pt]{0.350pt}{18.067pt}}
\put(620,483){\rule[-0.175pt]{4.818pt}{0.350pt}}
\put(620,558){\rule[-0.175pt]{4.818pt}{0.350pt}}
\put(781,375){\rule[-0.175pt]{0.350pt}{11.563pt}}
\put(771,375){\rule[-0.175pt]{4.818pt}{0.350pt}}
\put(771,423){\rule[-0.175pt]{4.818pt}{0.350pt}}
\put(1108,158){\rule[-0.175pt]{0.350pt}{6.022pt}}
\put(1098,158){\rule[-0.175pt]{4.818pt}{0.350pt}}
\put(1098,183){\rule[-0.175pt]{4.818pt}{0.350pt}}
\put(964,255){\rule[-0.175pt]{0.350pt}{9.636pt}}
\put(954,255){\rule[-0.175pt]{4.818pt}{0.350pt}}
\put(954,295){\rule[-0.175pt]{4.818pt}{0.350pt}}
\put(1134,158){\rule[-0.175pt]{0.350pt}{3.613pt}}
\put(1124,158){\rule[-0.175pt]{4.818pt}{0.350pt}}
\put(1124,173){\rule[-0.175pt]{4.818pt}{0.350pt}}
\put(1134,185){\rule[-0.175pt]{0.350pt}{4.095pt}}
\put(1124,185){\rule[-0.175pt]{4.818pt}{0.350pt}}
\put(1124,202){\rule[-0.175pt]{4.818pt}{0.350pt}}
\put(686,440){\rule[-0.175pt]{0.350pt}{19.031pt}}
\put(676,440){\rule[-0.175pt]{4.818pt}{0.350pt}}
\put(676,519){\rule[-0.175pt]{4.818pt}{0.350pt}}
\put(736,411){\rule[-0.175pt]{0.350pt}{9.154pt}}
\put(726,411){\rule[-0.175pt]{4.818pt}{0.350pt}}
\put(726,449){\rule[-0.175pt]{4.818pt}{0.350pt}}
\put(861,312){\rule[-0.175pt]{0.350pt}{8.191pt}}
\put(851,312){\rule[-0.175pt]{4.818pt}{0.350pt}}
\put(851,346){\rule[-0.175pt]{4.818pt}{0.350pt}}
\put(897,282){\rule[-0.175pt]{0.350pt}{14.454pt}}
\put(887,282){\rule[-0.175pt]{4.818pt}{0.350pt}}
\put(887,342){\rule[-0.175pt]{4.818pt}{0.350pt}}
\put(1025,228){\rule[-0.175pt]{0.350pt}{6.745pt}}
\put(1015,228){\rule[-0.175pt]{4.818pt}{0.350pt}}
\put(1015,256){\rule[-0.175pt]{4.818pt}{0.350pt}}
\put(1054,221){\rule[-0.175pt]{0.350pt}{8.191pt}}
\put(1044,221){\rule[-0.175pt]{4.818pt}{0.350pt}}
\put(1044,255){\rule[-0.175pt]{4.818pt}{0.350pt}}
\put(781,370){\rule[-0.175pt]{0.350pt}{8.913pt}}
\put(771,370){\rule[-0.175pt]{4.818pt}{0.350pt}}
\put(771,407){\rule[-0.175pt]{4.818pt}{0.350pt}}
\put(897,292){\rule[-0.175pt]{0.350pt}{5.782pt}}
\put(887,292){\rule[-0.175pt]{4.818pt}{0.350pt}}
\put(887,316){\rule[-0.175pt]{4.818pt}{0.350pt}}
\sbox{\plotpoint}{\rule[-0.250pt]{0.500pt}{0.500pt}}%
\put(1006,632){\makebox(0,0)[r]{quark model}}
\put(1028,632){\usebox{\plotpoint}}
\put(1048,632){\usebox{\plotpoint}}
\put(1069,632){\usebox{\plotpoint}}
\put(1090,632){\usebox{\plotpoint}}
\put(1094,632){\usebox{\plotpoint}}
\put(264,158){\usebox{\plotpoint}}
\put(264,158){\usebox{\plotpoint}}
\put(271,177){\usebox{\plotpoint}}
\put(279,196){\usebox{\plotpoint}}
\put(287,215){\usebox{\plotpoint}}
\put(295,234){\usebox{\plotpoint}}
\put(303,253){\usebox{\plotpoint}}
\put(311,273){\usebox{\plotpoint}}
\put(319,292){\usebox{\plotpoint}}
\put(328,311){\usebox{\plotpoint}}
\put(337,329){\usebox{\plotpoint}}
\put(346,348){\usebox{\plotpoint}}
\put(355,366){\usebox{\plotpoint}}
\put(366,385){\usebox{\plotpoint}}
\put(376,403){\usebox{\plotpoint}}
\put(386,421){\usebox{\plotpoint}}
\put(398,438){\usebox{\plotpoint}}
\put(410,455){\usebox{\plotpoint}}
\put(422,471){\usebox{\plotpoint}}
\put(436,487){\usebox{\plotpoint}}
\put(450,501){\usebox{\plotpoint}}
\put(467,514){\usebox{\plotpoint}}
\put(485,525){\usebox{\plotpoint}}
\put(503,534){\usebox{\plotpoint}}
\put(524,538){\usebox{\plotpoint}}
\put(544,540){\usebox{\plotpoint}}
\put(565,538){\usebox{\plotpoint}}
\put(585,533){\usebox{\plotpoint}}
\put(604,526){\usebox{\plotpoint}}
\put(623,517){\usebox{\plotpoint}}
\put(641,506){\usebox{\plotpoint}}
\put(659,496){\usebox{\plotpoint}}
\put(675,483){\usebox{\plotpoint}}
\put(692,471){\usebox{\plotpoint}}
\put(708,458){\usebox{\plotpoint}}
\put(724,445){\usebox{\plotpoint}}
\put(740,432){\usebox{\plotpoint}}
\put(756,418){\usebox{\plotpoint}}
\put(772,405){\usebox{\plotpoint}}
\put(788,392){\usebox{\plotpoint}}
\put(804,378){\usebox{\plotpoint}}
\put(820,365){\usebox{\plotpoint}}
\put(836,352){\usebox{\plotpoint}}
\put(853,340){\usebox{\plotpoint}}
\put(869,327){\usebox{\plotpoint}}
\put(886,315){\usebox{\plotpoint}}
\put(903,303){\usebox{\plotpoint}}
\put(920,291){\usebox{\plotpoint}}
\put(937,280){\usebox{\plotpoint}}
\put(955,269){\usebox{\plotpoint}}
\put(973,259){\usebox{\plotpoint}}
\put(992,249){\usebox{\plotpoint}}
\put(1011,241){\usebox{\plotpoint}}
\put(1030,232){\usebox{\plotpoint}}
\put(1049,225){\usebox{\plotpoint}}
\put(1068,218){\usebox{\plotpoint}}
\put(1088,210){\usebox{\plotpoint}}
\put(1108,205){\usebox{\plotpoint}}
\put(1128,199){\usebox{\plotpoint}}
\put(1136,197){\usebox{\plotpoint}}
\end{picture}
\caption{The radial wavefunctions for the $1S$ and $1P$ charmonium
computed using improved actions and two different lattice spacings.
Wavefunctions from a continuum quark model are also shown. Statistical
errors are negligible for the $1S$~wavefunction.}
\label{psi-wfcns}
\end{figure*}
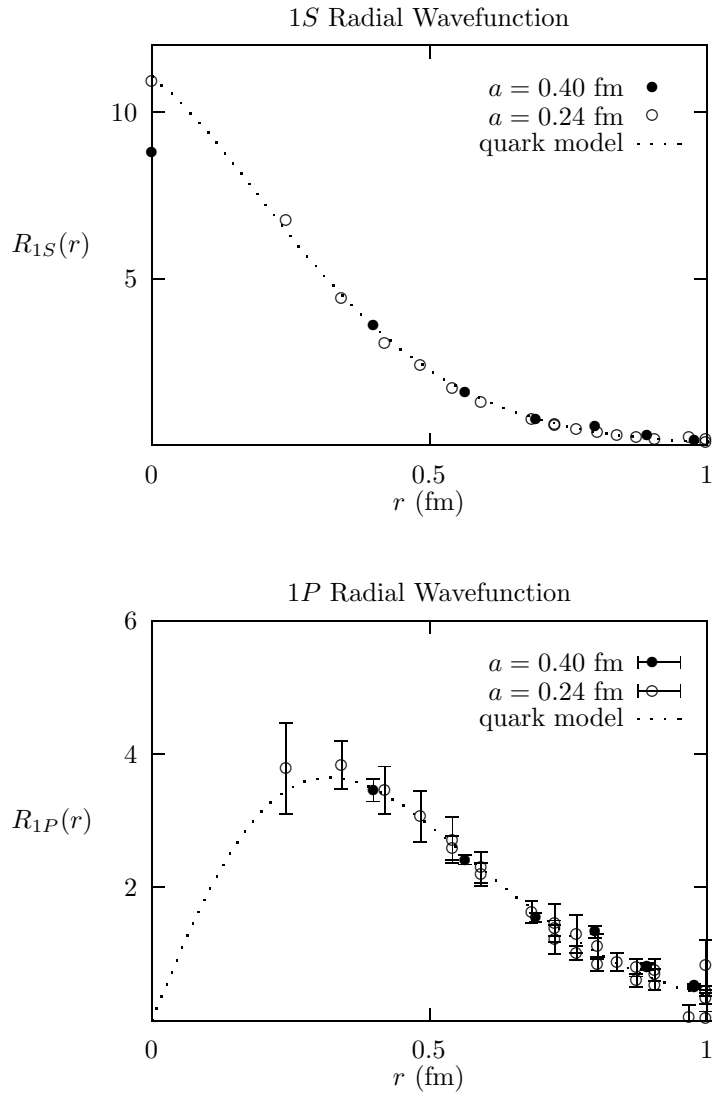

As final evidence of the quality of the improved actions, I show the
$1S$~and $1P$~radial wavefunctions for charmonium,
computed on lattices with two different
lattice spacings, using these actions (Figure~\ref{psi-wfcns}). The most
striking observation from these pictures is that the charge radius of
the~$\psi$ is almost exactly equal to one lattice spacing on the coarser
lattice, and yet the results from the coarser lattice are essentially
identical to those from the finer lattice. These data show that in general
one needs only a few lattice points per direction within a hadron to obtain
accurate results (few percent) from an $a^2$-accurate action.
Numerical experiments with the NRQCD action, where $a^2$~corrections are
easily turned on and off, suggest that
\be \label{nrqcd-errors}
r_{\rm hadron}\! \approx\! a  \Rightarrow
\cases{ \order(a^2)~\mbox{corrections}\approx \mbox{15--20\%} \cr
        \mbox{} \cr
        \order(\alpha_s a^2,a^4) \approx \mbox{2--4\%} }
\ee

\subsection{Gluons\,---\,Perfect Action}
DeGrand, Hasenfratz, Hasenfratz and Niedermeyer have recently computed the
classically perfect gluon action for QCD using analytic techniques for terms
quadratic and cubic in $A_\mu$ and numerical techniques for the rest of the
action\,\cite{hasen2}. This action is too complicated to use in
simulations but they have designed a much simpler, eight-parameter action that
approximates the perfect action to within a few percent for any field
configuration. This approximately perfect action is
built from the plaquette and parallelogram loops and powers of these. (The
powers are important.)

To test their action these authors computed the phase-transition
temperature~$T_c$ for lattice spacings  ranging from~$1/6T_c$ to~$1/2T_c$
using their action and using the Wilson action. They compare~$T_c$ with the
torelon mass~$\sigma$ (that is, the string tension on lattices of a fixed
spatial size). I show some of their results for $2\sqrt\sigma/T_c$ in
Table~\ref{tc-perfect}. Results for the Wilson action show finite-$a$ effects
of order 10\%; results from their improved action are $a$-independent to
within statistical errors of a couple percent. There is insufficient data
from other sources to know whether their results are the correct
$a\!=\!0$~results.
\begin{table}
\begin{tabular}{ccc}\hline
$N_t$ & Wilson & Perfect Action \\\hline
2 & 3.38 (4) & 3.09 (3) \\
3 & 3.15 (5) & 3.05 (5) \\
4 & 2.98 (3) & 3.08 (4)\\ \hline
\end{tabular}
\caption{Twice the ratio of the square root of the torelon
mass to the critical temperature for
the Wilson action and for the approximately perfect QCD action. The lattice
spacing is $1/N_tT_c$ and the torelons are for lattices with $2N_t$~sites in
the spatial direction.}
\label{tc-perfect}
\end{table}

\begin{figure*}
\hspace*{0.1in}
\setlength{\unitlength}{0.240900pt}
\ifx\plotpoint\undefined\newsavebox{\plotpoint}\fi
\sbox{\plotpoint}{\rule[-0.200pt]{0.400pt}{0.400pt}}%
\begin{picture}(1500,900)(0,0)
\font\gnuplot=cmr10 at 10pt
\gnuplot
\sbox{\plotpoint}{\rule[-0.200pt]{0.400pt}{0.400pt}}%
\put(220.0,476.0){\rule[-0.200pt]{292.934pt}{0.400pt}}
\put(220.0,113.0){\rule[-0.200pt]{0.400pt}{184.048pt}}
\put(220.0,285.0){\rule[-0.200pt]{4.818pt}{0.400pt}}
\put(198,285){\makebox(0,0)[r]{$-2$}}
\put(1416.0,285.0){\rule[-0.200pt]{4.818pt}{0.400pt}}
\put(220.0,476.0){\rule[-0.200pt]{4.818pt}{0.400pt}}
\put(198,476){\makebox(0,0)[r]{$0$}}
\put(1416.0,476.0){\rule[-0.200pt]{4.818pt}{0.400pt}}
\put(220.0,667.0){\rule[-0.200pt]{4.818pt}{0.400pt}}
\put(198,667){\makebox(0,0)[r]{$2$}}
\put(1416.0,667.0){\rule[-0.200pt]{4.818pt}{0.400pt}}
\put(220.0,858.0){\rule[-0.200pt]{4.818pt}{0.400pt}}
\put(198,858){\makebox(0,0)[r]{$4$}}
\put(1416.0,858.0){\rule[-0.200pt]{4.818pt}{0.400pt}}
\put(524.0,113.0){\rule[-0.200pt]{0.400pt}{4.818pt}}
\put(524,68){\makebox(0,0){$0.5$}}
\put(524.0,857.0){\rule[-0.200pt]{0.400pt}{4.818pt}}
\put(828.0,113.0){\rule[-0.200pt]{0.400pt}{4.818pt}}
\put(828,68){\makebox(0,0){$1$}}
\put(828.0,857.0){\rule[-0.200pt]{0.400pt}{4.818pt}}
\put(1132.0,113.0){\rule[-0.200pt]{0.400pt}{4.818pt}}
\put(1132,68){\makebox(0,0){$1.5$}}
\put(1132.0,857.0){\rule[-0.200pt]{0.400pt}{4.818pt}}
\put(220.0,113.0){\rule[-0.200pt]{292.934pt}{0.400pt}}
\put(1436.0,113.0){\rule[-0.200pt]{0.400pt}{184.048pt}}
\put(220.0,877.0){\rule[-0.200pt]{292.934pt}{0.400pt}}
\put(45,495){\makebox(0,0){$V(r)/T_c$}}
\put(828,23){\makebox(0,0){$r\,T_c$}}
\put(220.0,113.0){\rule[-0.200pt]{0.400pt}{184.048pt}}
\put(676,762){\makebox(0,0)[r]{$a\!\approx\!.36$~fm}}
\put(720,762){\circle{24}}
\put(524,335){\circle{24}}
\put(828,476){\circle{24}}
\put(1132,610){\circle{24}}
\put(650,411){\circle{24}}
\put(900,518){\circle{24}}
\put(1181,628){\circle{24}}
\put(1080,587){\circle{24}}
\put(1316,676){\circle{24}}
\put(747,466){\circle{24}}
\put(965,555){\circle{24}}
\put(1228,654){\circle{24}}
\put(1132,615){\circle{24}}
\put(1357,708){\circle{24}}
\put(1273,653){\circle{24}}
\put(676,717){\makebox(0,0)[r]{$a\!\approx\!.24$~fm}}
\put(720,717){\circle*{12}}
\put(372,207){\circle*{12}}
\put(524,324){\circle*{12}}
\put(676,404){\circle*{12}}
\put(828,476){\circle*{12}}
\put(980,516){\circle*{12}}
\put(1132,663){\circle*{12}}
\put(435,269){\circle*{12}}
\put(560,347){\circle*{12}}
\put(701,418){\circle*{12}}
\put(847,480){\circle*{12}}
\put(995,526){\circle*{12}}
\put(1145,607){\circle*{12}}
\put(650,396){\circle*{12}}
\put(768,450){\circle*{12}}
\put(900,507){\circle*{12}}
\put(1039,556){\circle*{12}}
\put(1181,632){\circle*{12}}
\put(865,488){\circle*{12}}
\put(980,542){\circle*{12}}
\put(1106,599){\circle*{12}}
\put(1240,634){\circle*{12}}
\put(1080,608){\circle*{12}}
\put(1193,627){\circle*{12}}
\put(1316,670){\circle*{12}}
\put(1295,684){\circle*{12}}
\put(1407,666){\circle*{12}}
\put(483,306){\circle*{12}}
\put(592,366){\circle*{12}}
\put(724,429){\circle*{12}}
\put(865,494){\circle*{12}}
\put(1010,546){\circle*{12}}
\put(1157,585){\circle*{12}}
\put(676,408){\circle*{12}}
\put(789,455){\circle*{12}}
\put(917,513){\circle*{12}}
\put(1053,566){\circle*{12}}
\put(1193,613){\circle*{12}}
\put(883,492){\circle*{12}}
\put(995,537){\circle*{12}}
\put(1119,580){\circle*{12}}
\put(1251,659){\circle*{12}}
\put(1093,591){\circle*{12}}
\put(1205,625){\circle*{12}}
\put(1327,718){\circle*{12}}
\put(1417,698){\circle*{12}}
\put(747,443){\circle*{12}}
\put(847,486){\circle*{12}}
\put(965,533){\circle*{12}}
\put(1093,584){\circle*{12}}
\put(1228,618){\circle*{12}}
\put(933,535){\circle*{12}}
\put(1039,547){\circle*{12}}
\put(1157,597){\circle*{12}}
\put(1284,707){\circle*{12}}
\put(1132,600){\circle*{12}}
\put(1240,665){\circle*{12}}
\put(1357,721){\circle*{12}}
\put(1010,535){\circle*{12}}
\put(1106,578){\circle*{12}}
\put(1217,666){\circle*{12}}
\put(1337,779){\circle*{12}}
\put(1193,621){\circle*{12}}
\put(1407,718){\circle*{12}}
\put(698.0,717.0){\rule[-0.200pt]{15.899pt}{0.400pt}}
\put(698.0,707.0){\rule[-0.200pt]{0.400pt}{4.818pt}}
\put(764.0,707.0){\rule[-0.200pt]{0.400pt}{4.818pt}}
\put(372,207){\usebox{\plotpoint}}
\put(362.0,207.0){\rule[-0.200pt]{4.818pt}{0.400pt}}
\put(362.0,207.0){\rule[-0.200pt]{4.818pt}{0.400pt}}
\put(524.0,323.0){\usebox{\plotpoint}}
\put(514.0,323.0){\rule[-0.200pt]{4.818pt}{0.400pt}}
\put(514.0,324.0){\rule[-0.200pt]{4.818pt}{0.400pt}}
\put(676.0,402.0){\rule[-0.200pt]{0.400pt}{0.723pt}}
\put(666.0,402.0){\rule[-0.200pt]{4.818pt}{0.400pt}}
\put(666.0,405.0){\rule[-0.200pt]{4.818pt}{0.400pt}}
\put(828.0,472.0){\rule[-0.200pt]{0.400pt}{1.927pt}}
\put(818.0,472.0){\rule[-0.200pt]{4.818pt}{0.400pt}}
\put(818.0,480.0){\rule[-0.200pt]{4.818pt}{0.400pt}}
\put(980.0,508.0){\rule[-0.200pt]{0.400pt}{3.854pt}}
\put(970.0,508.0){\rule[-0.200pt]{4.818pt}{0.400pt}}
\put(970.0,524.0){\rule[-0.200pt]{4.818pt}{0.400pt}}
\put(1132.0,608.0){\rule[-0.200pt]{0.400pt}{26.499pt}}
\put(1122.0,608.0){\rule[-0.200pt]{4.818pt}{0.400pt}}
\put(1122.0,718.0){\rule[-0.200pt]{4.818pt}{0.400pt}}
\put(435.0,268.0){\usebox{\plotpoint}}
\put(425.0,268.0){\rule[-0.200pt]{4.818pt}{0.400pt}}
\put(425.0,269.0){\rule[-0.200pt]{4.818pt}{0.400pt}}
\put(560.0,347.0){\usebox{\plotpoint}}
\put(550.0,347.0){\rule[-0.200pt]{4.818pt}{0.400pt}}
\put(550.0,348.0){\rule[-0.200pt]{4.818pt}{0.400pt}}
\put(701.0,417.0){\rule[-0.200pt]{0.400pt}{0.482pt}}
\put(691.0,417.0){\rule[-0.200pt]{4.818pt}{0.400pt}}
\put(691.0,419.0){\rule[-0.200pt]{4.818pt}{0.400pt}}
\put(847.0,477.0){\rule[-0.200pt]{0.400pt}{1.686pt}}
\put(837.0,477.0){\rule[-0.200pt]{4.818pt}{0.400pt}}
\put(837.0,484.0){\rule[-0.200pt]{4.818pt}{0.400pt}}
\put(995.0,519.0){\rule[-0.200pt]{0.400pt}{3.373pt}}
\put(985.0,519.0){\rule[-0.200pt]{4.818pt}{0.400pt}}
\put(985.0,533.0){\rule[-0.200pt]{4.818pt}{0.400pt}}
\put(1145.0,593.0){\rule[-0.200pt]{0.400pt}{6.745pt}}
\put(1135.0,593.0){\rule[-0.200pt]{4.818pt}{0.400pt}}
\put(1135.0,621.0){\rule[-0.200pt]{4.818pt}{0.400pt}}
\put(650.0,395.0){\rule[-0.200pt]{0.400pt}{0.482pt}}
\put(640.0,395.0){\rule[-0.200pt]{4.818pt}{0.400pt}}
\put(640.0,397.0){\rule[-0.200pt]{4.818pt}{0.400pt}}
\put(768.0,448.0){\rule[-0.200pt]{0.400pt}{0.964pt}}
\put(758.0,448.0){\rule[-0.200pt]{4.818pt}{0.400pt}}
\put(758.0,452.0){\rule[-0.200pt]{4.818pt}{0.400pt}}
\put(900.0,502.0){\rule[-0.200pt]{0.400pt}{2.650pt}}
\put(890.0,502.0){\rule[-0.200pt]{4.818pt}{0.400pt}}
\put(890.0,513.0){\rule[-0.200pt]{4.818pt}{0.400pt}}
\put(1039.0,546.0){\rule[-0.200pt]{0.400pt}{4.818pt}}
\put(1029.0,546.0){\rule[-0.200pt]{4.818pt}{0.400pt}}
\put(1029.0,566.0){\rule[-0.200pt]{4.818pt}{0.400pt}}
\put(1181.0,614.0){\rule[-0.200pt]{0.400pt}{8.672pt}}
\put(1171.0,614.0){\rule[-0.200pt]{4.818pt}{0.400pt}}
\put(1171.0,650.0){\rule[-0.200pt]{4.818pt}{0.400pt}}
\put(865.0,483.0){\rule[-0.200pt]{0.400pt}{2.409pt}}
\put(855.0,483.0){\rule[-0.200pt]{4.818pt}{0.400pt}}
\put(855.0,493.0){\rule[-0.200pt]{4.818pt}{0.400pt}}
\put(980.0,533.0){\rule[-0.200pt]{0.400pt}{4.336pt}}
\put(970.0,533.0){\rule[-0.200pt]{4.818pt}{0.400pt}}
\put(970.0,551.0){\rule[-0.200pt]{4.818pt}{0.400pt}}
\put(1106.0,578.0){\rule[-0.200pt]{0.400pt}{10.359pt}}
\put(1096.0,578.0){\rule[-0.200pt]{4.818pt}{0.400pt}}
\put(1096.0,621.0){\rule[-0.200pt]{4.818pt}{0.400pt}}
\put(1240.0,616.0){\rule[-0.200pt]{0.400pt}{8.913pt}}
\put(1230.0,616.0){\rule[-0.200pt]{4.818pt}{0.400pt}}
\put(1230.0,653.0){\rule[-0.200pt]{4.818pt}{0.400pt}}
\put(1080.0,576.0){\rule[-0.200pt]{0.400pt}{15.658pt}}
\put(1070.0,576.0){\rule[-0.200pt]{4.818pt}{0.400pt}}
\put(1070.0,641.0){\rule[-0.200pt]{4.818pt}{0.400pt}}
\put(1193.0,593.0){\rule[-0.200pt]{0.400pt}{16.622pt}}
\put(1183.0,593.0){\rule[-0.200pt]{4.818pt}{0.400pt}}
\put(1183.0,662.0){\rule[-0.200pt]{4.818pt}{0.400pt}}
\put(1316.0,636.0){\rule[-0.200pt]{0.400pt}{16.381pt}}
\put(1306.0,636.0){\rule[-0.200pt]{4.818pt}{0.400pt}}
\put(1306.0,704.0){\rule[-0.200pt]{4.818pt}{0.400pt}}
\put(1295.0,569.0){\rule[-0.200pt]{0.400pt}{55.407pt}}
\put(1285.0,569.0){\rule[-0.200pt]{4.818pt}{0.400pt}}
\put(1285.0,799.0){\rule[-0.200pt]{4.818pt}{0.400pt}}
\put(1407.0,633.0){\rule[-0.200pt]{0.400pt}{15.658pt}}
\put(1397.0,633.0){\rule[-0.200pt]{4.818pt}{0.400pt}}
\put(1397.0,698.0){\rule[-0.200pt]{4.818pt}{0.400pt}}
\put(483.0,305.0){\usebox{\plotpoint}}
\put(473.0,305.0){\rule[-0.200pt]{4.818pt}{0.400pt}}
\put(473.0,306.0){\rule[-0.200pt]{4.818pt}{0.400pt}}
\put(592.0,365.0){\rule[-0.200pt]{0.400pt}{0.482pt}}
\put(582.0,365.0){\rule[-0.200pt]{4.818pt}{0.400pt}}
\put(582.0,367.0){\rule[-0.200pt]{4.818pt}{0.400pt}}
\put(724.0,427.0){\rule[-0.200pt]{0.400pt}{0.964pt}}
\put(714.0,427.0){\rule[-0.200pt]{4.818pt}{0.400pt}}
\put(714.0,431.0){\rule[-0.200pt]{4.818pt}{0.400pt}}
\put(865.0,488.0){\rule[-0.200pt]{0.400pt}{2.891pt}}
\put(855.0,488.0){\rule[-0.200pt]{4.818pt}{0.400pt}}
\put(855.0,500.0){\rule[-0.200pt]{4.818pt}{0.400pt}}
\put(1010.0,533.0){\rule[-0.200pt]{0.400pt}{6.263pt}}
\put(1000.0,533.0){\rule[-0.200pt]{4.818pt}{0.400pt}}
\put(1000.0,559.0){\rule[-0.200pt]{4.818pt}{0.400pt}}
\put(1157.0,573.0){\rule[-0.200pt]{0.400pt}{5.541pt}}
\put(1147.0,573.0){\rule[-0.200pt]{4.818pt}{0.400pt}}
\put(1147.0,596.0){\rule[-0.200pt]{4.818pt}{0.400pt}}
\put(676.0,407.0){\rule[-0.200pt]{0.400pt}{0.723pt}}
\put(666.0,407.0){\rule[-0.200pt]{4.818pt}{0.400pt}}
\put(666.0,410.0){\rule[-0.200pt]{4.818pt}{0.400pt}}
\put(789.0,453.0){\rule[-0.200pt]{0.400pt}{1.204pt}}
\put(779.0,453.0){\rule[-0.200pt]{4.818pt}{0.400pt}}
\put(779.0,458.0){\rule[-0.200pt]{4.818pt}{0.400pt}}
\put(917.0,507.0){\rule[-0.200pt]{0.400pt}{2.650pt}}
\put(907.0,507.0){\rule[-0.200pt]{4.818pt}{0.400pt}}
\put(907.0,518.0){\rule[-0.200pt]{4.818pt}{0.400pt}}
\put(1053.0,553.0){\rule[-0.200pt]{0.400pt}{6.263pt}}
\put(1043.0,553.0){\rule[-0.200pt]{4.818pt}{0.400pt}}
\put(1043.0,579.0){\rule[-0.200pt]{4.818pt}{0.400pt}}
\put(1193.0,599.0){\rule[-0.200pt]{0.400pt}{6.504pt}}
\put(1183.0,599.0){\rule[-0.200pt]{4.818pt}{0.400pt}}
\put(1183.0,626.0){\rule[-0.200pt]{4.818pt}{0.400pt}}
\put(883.0,486.0){\rule[-0.200pt]{0.400pt}{2.891pt}}
\put(873.0,486.0){\rule[-0.200pt]{4.818pt}{0.400pt}}
\put(873.0,498.0){\rule[-0.200pt]{4.818pt}{0.400pt}}
\put(995.0,528.0){\rule[-0.200pt]{0.400pt}{4.095pt}}
\put(985.0,528.0){\rule[-0.200pt]{4.818pt}{0.400pt}}
\put(985.0,545.0){\rule[-0.200pt]{4.818pt}{0.400pt}}
\put(1119.0,565.0){\rule[-0.200pt]{0.400pt}{7.468pt}}
\put(1109.0,565.0){\rule[-0.200pt]{4.818pt}{0.400pt}}
\put(1109.0,596.0){\rule[-0.200pt]{4.818pt}{0.400pt}}
\put(1251.0,632.0){\rule[-0.200pt]{0.400pt}{13.009pt}}
\put(1241.0,632.0){\rule[-0.200pt]{4.818pt}{0.400pt}}
\put(1241.0,686.0){\rule[-0.200pt]{4.818pt}{0.400pt}}
\put(1093.0,565.0){\rule[-0.200pt]{0.400pt}{12.527pt}}
\put(1083.0,565.0){\rule[-0.200pt]{4.818pt}{0.400pt}}
\put(1083.0,617.0){\rule[-0.200pt]{4.818pt}{0.400pt}}
\put(1205.0,592.0){\rule[-0.200pt]{0.400pt}{16.140pt}}
\put(1195.0,592.0){\rule[-0.200pt]{4.818pt}{0.400pt}}
\put(1195.0,659.0){\rule[-0.200pt]{4.818pt}{0.400pt}}
\put(1327.0,650.0){\rule[-0.200pt]{0.400pt}{32.762pt}}
\put(1317.0,650.0){\rule[-0.200pt]{4.818pt}{0.400pt}}
\put(1317.0,786.0){\rule[-0.200pt]{4.818pt}{0.400pt}}
\put(1417.0,644.0){\rule[-0.200pt]{0.400pt}{26.258pt}}
\put(1407.0,644.0){\rule[-0.200pt]{4.818pt}{0.400pt}}
\put(1407.0,753.0){\rule[-0.200pt]{4.818pt}{0.400pt}}
\put(747.0,439.0){\rule[-0.200pt]{0.400pt}{1.927pt}}
\put(737.0,439.0){\rule[-0.200pt]{4.818pt}{0.400pt}}
\put(737.0,447.0){\rule[-0.200pt]{4.818pt}{0.400pt}}
\put(847.0,480.0){\rule[-0.200pt]{0.400pt}{2.650pt}}
\put(837.0,480.0){\rule[-0.200pt]{4.818pt}{0.400pt}}
\put(837.0,491.0){\rule[-0.200pt]{4.818pt}{0.400pt}}
\put(965.0,523.0){\rule[-0.200pt]{0.400pt}{4.577pt}}
\put(955.0,523.0){\rule[-0.200pt]{4.818pt}{0.400pt}}
\put(955.0,542.0){\rule[-0.200pt]{4.818pt}{0.400pt}}
\put(1093.0,561.0){\rule[-0.200pt]{0.400pt}{11.322pt}}
\put(1083.0,561.0){\rule[-0.200pt]{4.818pt}{0.400pt}}
\put(1083.0,608.0){\rule[-0.200pt]{4.818pt}{0.400pt}}
\put(1228.0,598.0){\rule[-0.200pt]{0.400pt}{9.636pt}}
\put(1218.0,598.0){\rule[-0.200pt]{4.818pt}{0.400pt}}
\put(1218.0,638.0){\rule[-0.200pt]{4.818pt}{0.400pt}}
\put(933.0,522.0){\rule[-0.200pt]{0.400pt}{6.022pt}}
\put(923.0,522.0){\rule[-0.200pt]{4.818pt}{0.400pt}}
\put(923.0,547.0){\rule[-0.200pt]{4.818pt}{0.400pt}}
\put(1039.0,536.0){\rule[-0.200pt]{0.400pt}{5.300pt}}
\put(1029.0,536.0){\rule[-0.200pt]{4.818pt}{0.400pt}}
\put(1029.0,558.0){\rule[-0.200pt]{4.818pt}{0.400pt}}
\put(1157.0,574.0){\rule[-0.200pt]{0.400pt}{10.840pt}}
\put(1147.0,574.0){\rule[-0.200pt]{4.818pt}{0.400pt}}
\put(1147.0,619.0){\rule[-0.200pt]{4.818pt}{0.400pt}}
\put(1284.0,644.0){\rule[-0.200pt]{0.400pt}{30.594pt}}
\put(1274.0,644.0){\rule[-0.200pt]{4.818pt}{0.400pt}}
\put(1274.0,771.0){\rule[-0.200pt]{4.818pt}{0.400pt}}
\put(1132.0,571.0){\rule[-0.200pt]{0.400pt}{14.213pt}}
\put(1122.0,571.0){\rule[-0.200pt]{4.818pt}{0.400pt}}
\put(1122.0,630.0){\rule[-0.200pt]{4.818pt}{0.400pt}}
\put(1240.0,602.0){\rule[-0.200pt]{0.400pt}{30.112pt}}
\put(1230.0,602.0){\rule[-0.200pt]{4.818pt}{0.400pt}}
\put(1230.0,727.0){\rule[-0.200pt]{4.818pt}{0.400pt}}
\put(1357.0,651.0){\rule[-0.200pt]{0.400pt}{33.967pt}}
\put(1347.0,651.0){\rule[-0.200pt]{4.818pt}{0.400pt}}
\put(1347.0,792.0){\rule[-0.200pt]{4.818pt}{0.400pt}}
\put(1010.0,516.0){\rule[-0.200pt]{0.400pt}{8.913pt}}
\put(1000.0,516.0){\rule[-0.200pt]{4.818pt}{0.400pt}}
\put(1000.0,553.0){\rule[-0.200pt]{4.818pt}{0.400pt}}
\put(1106.0,554.0){\rule[-0.200pt]{0.400pt}{11.563pt}}
\put(1096.0,554.0){\rule[-0.200pt]{4.818pt}{0.400pt}}
\put(1096.0,602.0){\rule[-0.200pt]{4.818pt}{0.400pt}}
\put(1217.0,585.0){\rule[-0.200pt]{0.400pt}{39.267pt}}
\put(1207.0,585.0){\rule[-0.200pt]{4.818pt}{0.400pt}}
\put(1207.0,748.0){\rule[-0.200pt]{4.818pt}{0.400pt}}
\put(1337.0,546.0){\rule[-0.200pt]{0.400pt}{79.738pt}}
\put(1327.0,546.0){\rule[-0.200pt]{4.818pt}{0.400pt}}
\put(1327.0,877.0){\rule[-0.200pt]{4.818pt}{0.400pt}}
\put(1193.0,575.0){\rule[-0.200pt]{0.400pt}{22.163pt}}
\put(1183.0,575.0){\rule[-0.200pt]{4.818pt}{0.400pt}}
\put(1183.0,667.0){\rule[-0.200pt]{4.818pt}{0.400pt}}
\put(1407.0,645.0){\rule[-0.200pt]{0.400pt}{35.171pt}}
\put(1397.0,645.0){\rule[-0.200pt]{4.818pt}{0.400pt}}
\put(1397.0,791.0){\rule[-0.200pt]{4.818pt}{0.400pt}}
\end{picture}
\caption{The static quark potential computed using an approximately perfect
gluon action for two different different lattice spacings.}
\label{potl-pa}
\end{figure*}
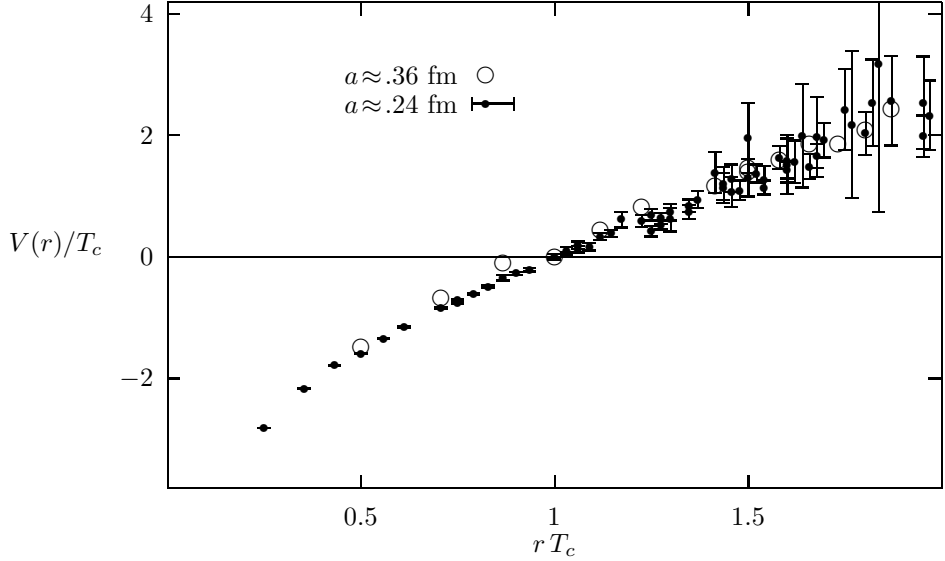

They also computed the static-quark potential for two different lattice
spacings. The results, in Figure~\ref{potl-pa}, show  large-$r$
behavior that is independent of the lattice spacing. At smaller radii, results
from the coarser lattice show some failure of rotational invariance. These
finite-$a$ errors are roughly twice as large as those we find with the
Symanzik-improved gluon action of the previous section. DeGrand and
collaborators suggest that the difference is because they used the blocked
gluon operator to compute the potential, which introduces
additional finite-$a$ errors. I am not certain this is  the explanation since
the $a^2$~errors in their measurement are rotationally
invariant, and the $a^4$~errors ought to be pretty small.
Another possibility is that the errors in the potential are
due to tadpole effects. The finite-$a$ errors in their potential are almost
identical to what we obtain from the Symanzik-improved action if we omit
tadpole improvement (with or without $\alpha_s$~corrections). This
suggests that a tadpole-improved version of the perfect action should be
tried.

\subsection{Light Quarks}
All standard relativistic quark actions are built using a first-order
centered difference operator,
\be
\Delta^{(1)}\psi \equiv \frac{\psi(x+a)-\psi(x-a)}{2a}
\ee
(ignoring gauge fields). Consequently the effective lattice spacing in
light-quark actions is $a_{\rm eff} \!=\!2a$, and not $a_{\rm eff} \!=\!a$ as
in gluon and heavy-quark (NRQCD) actions. This makes the light-quark actions
the least accurate component in lattice simulations and the most in  need of
repair. Based on our experience with NRQCD (\eq{nrqcd-errors}), we expect
order~$a^2$ corrections to a light-quark action to be of order~20\% when
$a\!=\!.4$~fm, which corresponds to an $a_{\rm eff}$ of order the
light-hardron radius. To get errors down to a few percent with such a coarse
lattice requires an action corrected for $\order(a^2)$~effects and possibly
also for~$\order(a^3)$.

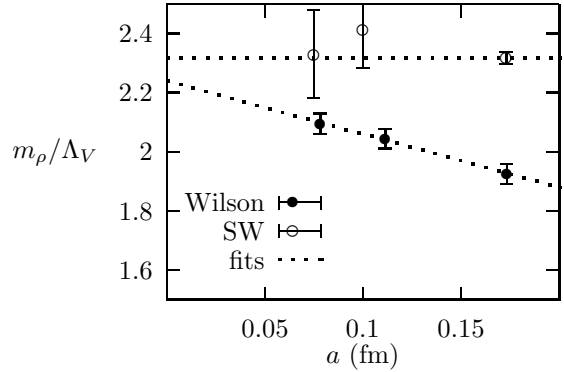
\begin{figure}
\setlength{\unitlength}{0.240900pt}
\ifx\plotpoint\undefined\newsavebox{\plotpoint}\fi
\sbox{\plotpoint}{\rule[-0.200pt]{0.400pt}{0.400pt}}%
\begin{picture}(900,600)(0,0)
\font\gnuplot=cmr10 at 10pt
\gnuplot
\sbox{\plotpoint}{\rule[-0.200pt]{0.400pt}{0.400pt}}%
\put(220.0,113.0){\rule[-0.200pt]{0.400pt}{111.778pt}}
\put(220.0,159.0){\rule[-0.200pt]{4.818pt}{0.400pt}}
\put(198,159){\makebox(0,0)[r]{$1.6$}}
\put(816.0,159.0){\rule[-0.200pt]{4.818pt}{0.400pt}}
\put(220.0,252.0){\rule[-0.200pt]{4.818pt}{0.400pt}}
\put(198,252){\makebox(0,0)[r]{$1.8$}}
\put(816.0,252.0){\rule[-0.200pt]{4.818pt}{0.400pt}}
\put(220.0,345.0){\rule[-0.200pt]{4.818pt}{0.400pt}}
\put(198,345){\makebox(0,0)[r]{$2$}}
\put(816.0,345.0){\rule[-0.200pt]{4.818pt}{0.400pt}}
\put(220.0,438.0){\rule[-0.200pt]{4.818pt}{0.400pt}}
\put(198,438){\makebox(0,0)[r]{$2.2$}}
\put(816.0,438.0){\rule[-0.200pt]{4.818pt}{0.400pt}}
\put(220.0,531.0){\rule[-0.200pt]{4.818pt}{0.400pt}}
\put(198,531){\makebox(0,0)[r]{$2.4$}}
\put(816.0,531.0){\rule[-0.200pt]{4.818pt}{0.400pt}}
\put(374.0,113.0){\rule[-0.200pt]{0.400pt}{4.818pt}}
\put(374,68){\makebox(0,0){$0.05$}}
\put(374.0,557.0){\rule[-0.200pt]{0.400pt}{4.818pt}}
\put(528.0,113.0){\rule[-0.200pt]{0.400pt}{4.818pt}}
\put(528,68){\makebox(0,0){$0.1$}}
\put(528.0,557.0){\rule[-0.200pt]{0.400pt}{4.818pt}}
\put(682.0,113.0){\rule[-0.200pt]{0.400pt}{4.818pt}}
\put(682,68){\makebox(0,0){$0.15$}}
\put(682.0,557.0){\rule[-0.200pt]{0.400pt}{4.818pt}}
\put(220.0,113.0){\rule[-0.200pt]{148.394pt}{0.400pt}}
\put(836.0,113.0){\rule[-0.200pt]{0.400pt}{111.778pt}}
\put(220.0,577.0){\rule[-0.200pt]{148.394pt}{0.400pt}}
\put(45,345){\makebox(0,0){$m_\rho/\Lambda_V$}}
\put(528,23){\makebox(0,0){$a$ (fm)}}
\put(220.0,113.0){\rule[-0.200pt]{0.400pt}{111.778pt}}
\put(374,266){\makebox(0,0)[r]{Wilson}}
\put(418,266){\circle*{18}}
\put(754,310){\circle*{18}}
\put(563,366){\circle*{18}}
\put(461,389){\circle*{18}}
\put(396.0,266.0){\rule[-0.200pt]{15.899pt}{0.400pt}}
\put(396.0,256.0){\rule[-0.200pt]{0.400pt}{4.818pt}}
\put(462.0,256.0){\rule[-0.200pt]{0.400pt}{4.818pt}}
\put(754.0,294.0){\rule[-0.200pt]{0.400pt}{7.709pt}}
\put(744.0,294.0){\rule[-0.200pt]{4.818pt}{0.400pt}}
\put(744.0,326.0){\rule[-0.200pt]{4.818pt}{0.400pt}}
\put(563.0,350.0){\rule[-0.200pt]{0.400pt}{7.468pt}}
\put(553.0,350.0){\rule[-0.200pt]{4.818pt}{0.400pt}}
\put(553.0,381.0){\rule[-0.200pt]{4.818pt}{0.400pt}}
\put(461.0,373.0){\rule[-0.200pt]{0.400pt}{7.709pt}}
\put(451.0,373.0){\rule[-0.200pt]{4.818pt}{0.400pt}}
\put(451.0,405.0){\rule[-0.200pt]{4.818pt}{0.400pt}}
\put(374,221){\makebox(0,0)[r]{SW}}
\put(418,221){\circle{18}}
\put(753,493){\circle{18}}
\put(528,537){\circle{18}}
\put(451,498){\circle{18}}
\put(396.0,221.0){\rule[-0.200pt]{15.899pt}{0.400pt}}
\put(396.0,211.0){\rule[-0.200pt]{0.400pt}{4.818pt}}
\put(462.0,211.0){\rule[-0.200pt]{0.400pt}{4.818pt}}
\put(753.0,483.0){\rule[-0.200pt]{0.400pt}{4.577pt}}
\put(743.0,483.0){\rule[-0.200pt]{4.818pt}{0.400pt}}
\put(743.0,502.0){\rule[-0.200pt]{4.818pt}{0.400pt}}
\put(528.0,477.0){\rule[-0.200pt]{0.400pt}{24.090pt}}
\put(518.0,477.0){\rule[-0.200pt]{4.818pt}{0.400pt}}
\put(518.0,577.0){\rule[-0.200pt]{4.818pt}{0.400pt}}
\put(451.0,429.0){\rule[-0.200pt]{0.400pt}{33.485pt}}
\put(441.0,429.0){\rule[-0.200pt]{4.818pt}{0.400pt}}
\put(441.0,568.0){\rule[-0.200pt]{4.818pt}{0.400pt}}
\sbox{\plotpoint}{\rule[-0.500pt]{1.000pt}{1.000pt}}%
\put(374,176){\makebox(0,0)[r]{fits}}
\multiput(396,176)(20.756,0.000){4}{\usebox{\plotpoint}}
\put(462,176){\usebox{\plotpoint}}
\put(220,456){\usebox{\plotpoint}}
\put(220.00,456.00){\usebox{\plotpoint}}
\multiput(226,455)(19.690,-6.563){0}{\usebox{\plotpoint}}
\multiput(232,453)(19.957,-5.702){0}{\usebox{\plotpoint}}
\put(240.05,450.82){\usebox{\plotpoint}}
\multiput(245,450)(19.690,-6.563){0}{\usebox{\plotpoint}}
\multiput(251,448)(19.690,-6.563){0}{\usebox{\plotpoint}}
\put(260.06,445.56){\usebox{\plotpoint}}
\multiput(264,445)(19.690,-6.563){0}{\usebox{\plotpoint}}
\multiput(270,443)(19.690,-6.563){0}{\usebox{\plotpoint}}
\put(279.92,439.69){\usebox{\plotpoint}}
\multiput(282,439)(20.473,-3.412){0}{\usebox{\plotpoint}}
\multiput(288,438)(19.957,-5.702){0}{\usebox{\plotpoint}}
\put(299.93,434.36){\usebox{\plotpoint}}
\multiput(301,434)(20.473,-3.412){0}{\usebox{\plotpoint}}
\multiput(307,433)(19.690,-6.563){0}{\usebox{\plotpoint}}
\put(319.94,429.02){\usebox{\plotpoint}}
\multiput(320,429)(20.473,-3.412){0}{\usebox{\plotpoint}}
\multiput(326,428)(19.690,-6.563){0}{\usebox{\plotpoint}}
\multiput(332,426)(19.690,-6.563){0}{\usebox{\plotpoint}}
\put(339.94,423.68){\usebox{\plotpoint}}
\multiput(344,423)(19.957,-5.702){0}{\usebox{\plotpoint}}
\multiput(351,421)(19.690,-6.563){0}{\usebox{\plotpoint}}
\put(359.87,418.04){\usebox{\plotpoint}}
\multiput(363,417)(20.473,-3.412){0}{\usebox{\plotpoint}}
\multiput(369,416)(19.957,-5.702){0}{\usebox{\plotpoint}}
\put(379.89,412.70){\usebox{\plotpoint}}
\multiput(382,412)(20.473,-3.412){0}{\usebox{\plotpoint}}
\multiput(388,411)(19.690,-6.563){0}{\usebox{\plotpoint}}
\put(399.81,407.06){\usebox{\plotpoint}}
\multiput(400,407)(20.547,-2.935){0}{\usebox{\plotpoint}}
\multiput(407,406)(19.690,-6.563){0}{\usebox{\plotpoint}}
\multiput(413,404)(19.690,-6.563){0}{\usebox{\plotpoint}}
\put(419.82,401.86){\usebox{\plotpoint}}
\multiput(425,401)(19.957,-5.702){0}{\usebox{\plotpoint}}
\multiput(432,399)(19.690,-6.563){0}{\usebox{\plotpoint}}
\put(439.80,396.40){\usebox{\plotpoint}}
\multiput(444,395)(20.473,-3.412){0}{\usebox{\plotpoint}}
\multiput(450,394)(19.690,-6.563){0}{\usebox{\plotpoint}}
\put(459.77,390.92){\usebox{\plotpoint}}
\multiput(463,390)(20.473,-3.412){0}{\usebox{\plotpoint}}
\multiput(469,389)(19.690,-6.563){0}{\usebox{\plotpoint}}
\put(479.74,385.42){\usebox{\plotpoint}}
\multiput(481,385)(20.547,-2.935){0}{\usebox{\plotpoint}}
\multiput(488,384)(19.690,-6.563){0}{\usebox{\plotpoint}}
\put(499.72,380.09){\usebox{\plotpoint}}
\multiput(500,380)(20.473,-3.412){0}{\usebox{\plotpoint}}
\multiput(506,379)(19.690,-6.563){0}{\usebox{\plotpoint}}
\multiput(512,377)(19.957,-5.702){0}{\usebox{\plotpoint}}
\put(519.73,374.76){\usebox{\plotpoint}}
\multiput(525,373)(20.473,-3.412){0}{\usebox{\plotpoint}}
\multiput(531,372)(19.690,-6.563){0}{\usebox{\plotpoint}}
\put(539.69,369.23){\usebox{\plotpoint}}
\multiput(544,368)(20.473,-3.412){0}{\usebox{\plotpoint}}
\multiput(550,367)(19.690,-6.563){0}{\usebox{\plotpoint}}
\put(559.66,363.78){\usebox{\plotpoint}}
\multiput(562,363)(20.473,-3.412){0}{\usebox{\plotpoint}}
\multiput(568,362)(19.957,-5.702){0}{\usebox{\plotpoint}}
\put(579.68,358.44){\usebox{\plotpoint}}
\multiput(581,358)(20.473,-3.412){0}{\usebox{\plotpoint}}
\multiput(587,357)(19.690,-6.563){0}{\usebox{\plotpoint}}
\put(599.69,353.09){\usebox{\plotpoint}}
\multiput(600,353)(19.690,-6.563){0}{\usebox{\plotpoint}}
\multiput(606,351)(20.473,-3.412){0}{\usebox{\plotpoint}}
\multiput(612,350)(19.690,-6.563){0}{\usebox{\plotpoint}}
\put(619.61,347.46){\usebox{\plotpoint}}
\multiput(624,346)(20.547,-2.935){0}{\usebox{\plotpoint}}
\multiput(631,345)(19.690,-6.563){0}{\usebox{\plotpoint}}
\put(639.59,342.14){\usebox{\plotpoint}}
\multiput(643,341)(20.473,-3.412){0}{\usebox{\plotpoint}}
\multiput(649,340)(19.957,-5.702){0}{\usebox{\plotpoint}}
\put(659.61,336.80){\usebox{\plotpoint}}
\multiput(662,336)(20.473,-3.412){0}{\usebox{\plotpoint}}
\multiput(668,335)(19.690,-6.563){0}{\usebox{\plotpoint}}
\put(679.53,331.16){\usebox{\plotpoint}}
\multiput(680,331)(19.957,-5.702){0}{\usebox{\plotpoint}}
\multiput(687,329)(20.473,-3.412){0}{\usebox{\plotpoint}}
\multiput(693,328)(19.690,-6.563){0}{\usebox{\plotpoint}}
\put(699.54,325.82){\usebox{\plotpoint}}
\multiput(705,324)(20.547,-2.935){0}{\usebox{\plotpoint}}
\multiput(712,323)(19.690,-6.563){0}{\usebox{\plotpoint}}
\put(719.52,320.49){\usebox{\plotpoint}}
\multiput(724,319)(20.473,-3.412){0}{\usebox{\plotpoint}}
\multiput(730,318)(19.690,-6.563){0}{\usebox{\plotpoint}}
\put(739.49,315.00){\usebox{\plotpoint}}
\multiput(743,314)(20.473,-3.412){0}{\usebox{\plotpoint}}
\multiput(749,313)(19.690,-6.563){0}{\usebox{\plotpoint}}
\put(759.45,309.52){\usebox{\plotpoint}}
\multiput(761,309)(19.957,-5.702){0}{\usebox{\plotpoint}}
\multiput(768,307)(20.473,-3.412){0}{\usebox{\plotpoint}}
\put(779.47,304.18){\usebox{\plotpoint}}
\multiput(780,304)(19.690,-6.563){0}{\usebox{\plotpoint}}
\multiput(786,302)(20.473,-3.412){0}{\usebox{\plotpoint}}
\multiput(792,301)(19.957,-5.702){0}{\usebox{\plotpoint}}
\put(799.48,298.84){\usebox{\plotpoint}}
\multiput(805,297)(20.473,-3.412){0}{\usebox{\plotpoint}}
\multiput(811,296)(19.690,-6.563){0}{\usebox{\plotpoint}}
\put(819.43,293.30){\usebox{\plotpoint}}
\multiput(824,292)(20.473,-3.412){0}{\usebox{\plotpoint}}
\multiput(830,291)(19.690,-6.563){0}{\usebox{\plotpoint}}
\put(836,289){\usebox{\plotpoint}}
\put(220,493){\usebox{\plotpoint}}
\put(220.00,493.00){\usebox{\plotpoint}}
\multiput(226,493)(20.756,0.000){0}{\usebox{\plotpoint}}
\multiput(232,493)(20.756,0.000){0}{\usebox{\plotpoint}}
\put(240.76,493.00){\usebox{\plotpoint}}
\multiput(245,493)(20.756,0.000){0}{\usebox{\plotpoint}}
\multiput(251,493)(20.756,0.000){0}{\usebox{\plotpoint}}
\put(261.51,493.00){\usebox{\plotpoint}}
\multiput(264,493)(20.756,0.000){0}{\usebox{\plotpoint}}
\multiput(270,493)(20.756,0.000){0}{\usebox{\plotpoint}}
\multiput(276,493)(20.756,0.000){0}{\usebox{\plotpoint}}
\put(282.27,493.00){\usebox{\plotpoint}}
\multiput(288,493)(20.756,0.000){0}{\usebox{\plotpoint}}
\multiput(295,493)(20.756,0.000){0}{\usebox{\plotpoint}}
\put(303.02,493.00){\usebox{\plotpoint}}
\multiput(307,493)(20.756,0.000){0}{\usebox{\plotpoint}}
\multiput(313,493)(20.756,0.000){0}{\usebox{\plotpoint}}
\put(323.78,493.00){\usebox{\plotpoint}}
\multiput(326,493)(20.756,0.000){0}{\usebox{\plotpoint}}
\multiput(332,493)(20.756,0.000){0}{\usebox{\plotpoint}}
\multiput(338,493)(20.756,0.000){0}{\usebox{\plotpoint}}
\put(344.53,493.00){\usebox{\plotpoint}}
\multiput(351,493)(20.756,0.000){0}{\usebox{\plotpoint}}
\multiput(357,493)(20.756,0.000){0}{\usebox{\plotpoint}}
\put(365.29,493.00){\usebox{\plotpoint}}
\multiput(369,493)(20.756,0.000){0}{\usebox{\plotpoint}}
\multiput(376,493)(20.756,0.000){0}{\usebox{\plotpoint}}
\put(386.04,493.00){\usebox{\plotpoint}}
\multiput(388,493)(20.756,0.000){0}{\usebox{\plotpoint}}
\multiput(394,493)(20.756,0.000){0}{\usebox{\plotpoint}}
\put(406.80,493.00){\usebox{\plotpoint}}
\multiput(407,493)(20.756,0.000){0}{\usebox{\plotpoint}}
\multiput(413,493)(20.756,0.000){0}{\usebox{\plotpoint}}
\multiput(419,493)(20.756,0.000){0}{\usebox{\plotpoint}}
\put(427.55,493.00){\usebox{\plotpoint}}
\multiput(432,493)(20.756,0.000){0}{\usebox{\plotpoint}}
\multiput(438,493)(20.756,0.000){0}{\usebox{\plotpoint}}
\put(448.31,493.00){\usebox{\plotpoint}}
\multiput(450,493)(20.756,0.000){0}{\usebox{\plotpoint}}
\multiput(456,493)(20.756,0.000){0}{\usebox{\plotpoint}}
\multiput(463,493)(20.756,0.000){0}{\usebox{\plotpoint}}
\put(469.07,493.00){\usebox{\plotpoint}}
\multiput(475,493)(20.756,0.000){0}{\usebox{\plotpoint}}
\multiput(481,493)(20.756,0.000){0}{\usebox{\plotpoint}}
\put(489.82,493.00){\usebox{\plotpoint}}
\multiput(494,493)(20.756,0.000){0}{\usebox{\plotpoint}}
\multiput(500,493)(20.756,0.000){0}{\usebox{\plotpoint}}
\put(510.58,493.00){\usebox{\plotpoint}}
\multiput(512,493)(20.756,0.000){0}{\usebox{\plotpoint}}
\multiput(519,493)(20.756,0.000){0}{\usebox{\plotpoint}}
\multiput(525,493)(20.756,0.000){0}{\usebox{\plotpoint}}
\put(531.33,493.00){\usebox{\plotpoint}}
\multiput(537,493)(20.756,0.000){0}{\usebox{\plotpoint}}
\multiput(544,493)(20.756,0.000){0}{\usebox{\plotpoint}}
\put(552.09,493.00){\usebox{\plotpoint}}
\multiput(556,493)(20.756,0.000){0}{\usebox{\plotpoint}}
\multiput(562,493)(20.756,0.000){0}{\usebox{\plotpoint}}
\put(572.84,493.00){\usebox{\plotpoint}}
\multiput(575,493)(20.756,0.000){0}{\usebox{\plotpoint}}
\multiput(581,493)(20.756,0.000){0}{\usebox{\plotpoint}}
\multiput(587,493)(20.756,0.000){0}{\usebox{\plotpoint}}
\put(593.60,493.00){\usebox{\plotpoint}}
\multiput(600,493)(20.756,0.000){0}{\usebox{\plotpoint}}
\multiput(606,493)(20.756,0.000){0}{\usebox{\plotpoint}}
\put(614.35,493.00){\usebox{\plotpoint}}
\multiput(618,493)(20.756,0.000){0}{\usebox{\plotpoint}}
\multiput(624,493)(20.756,0.000){0}{\usebox{\plotpoint}}
\put(635.11,493.00){\usebox{\plotpoint}}
\multiput(637,493)(20.756,0.000){0}{\usebox{\plotpoint}}
\multiput(643,493)(20.756,0.000){0}{\usebox{\plotpoint}}
\put(655.87,493.00){\usebox{\plotpoint}}
\multiput(656,493)(20.756,0.000){0}{\usebox{\plotpoint}}
\multiput(662,493)(20.756,0.000){0}{\usebox{\plotpoint}}
\multiput(668,493)(20.756,0.000){0}{\usebox{\plotpoint}}
\put(676.62,493.00){\usebox{\plotpoint}}
\multiput(680,493)(20.756,0.000){0}{\usebox{\plotpoint}}
\multiput(687,493)(20.756,0.000){0}{\usebox{\plotpoint}}
\put(697.38,493.00){\usebox{\plotpoint}}
\multiput(699,493)(20.756,0.000){0}{\usebox{\plotpoint}}
\multiput(705,493)(20.756,0.000){0}{\usebox{\plotpoint}}
\multiput(712,493)(20.756,0.000){0}{\usebox{\plotpoint}}
\put(718.13,493.00){\usebox{\plotpoint}}
\multiput(724,493)(20.756,0.000){0}{\usebox{\plotpoint}}
\multiput(730,493)(20.756,0.000){0}{\usebox{\plotpoint}}
\put(738.89,493.00){\usebox{\plotpoint}}
\multiput(743,493)(20.756,0.000){0}{\usebox{\plotpoint}}
\multiput(749,493)(20.756,0.000){0}{\usebox{\plotpoint}}
\put(759.64,493.00){\usebox{\plotpoint}}
\multiput(761,493)(20.756,0.000){0}{\usebox{\plotpoint}}
\multiput(768,493)(20.756,0.000){0}{\usebox{\plotpoint}}
\multiput(774,493)(20.756,0.000){0}{\usebox{\plotpoint}}
\put(780.40,493.00){\usebox{\plotpoint}}
\multiput(786,493)(20.756,0.000){0}{\usebox{\plotpoint}}
\multiput(792,493)(20.756,0.000){0}{\usebox{\plotpoint}}
\put(801.15,493.00){\usebox{\plotpoint}}
\multiput(805,493)(20.756,0.000){0}{\usebox{\plotpoint}}
\multiput(811,493)(20.756,0.000){0}{\usebox{\plotpoint}}
\put(821.91,493.00){\usebox{\plotpoint}}
\multiput(824,493)(20.756,0.000){0}{\usebox{\plotpoint}}
\multiput(830,493)(20.756,0.000){0}{\usebox{\plotpoint}}
\put(836,493){\usebox{\plotpoint}}
\end{picture}
\caption{The $\rho$~mass divided by the~$\alpha_V$ scale
parameter~$\Lambda_V$ versus $a$. Results are shown for the Wilson and SW quark
actions. The results are for quenched simulations with $\beta$'s ranging
from~5.7 to~6.2; the lattice spacing is assumed to be .1~fm at $\beta\!=\!6$.}
\label{rho-sw}
\end{figure}

The most accurate light-quark action in common use today is the
tadpole-improved version of the Sheikholeslami-Wohlert (SW) action:
\bearray
S_{\rm SW} &=& \sum \overline\psi\left\{-\Delta^{(1)}\cdot\gamma + m
\right.\nl  && \left. -\frac{ra}{2}\left(\Delta^{(2)} +
\frac{g\sigma\cdot F}{2}\right)\right\}\psi,
\eearray
where parameter $r\!=\!1$ is usually chosen. There is now considerable
evidence showing that this action, which has $\order(a^2)$~errors, is much
more accurate than the Wilson action for quarks, which has
$\order(a)$~errors. At this conference Brian Gough presented data from
Fermilab on quark masses that show a significant reduction in finite-$a$
errors with the SW~action. Also Hugh Shanahan presented new  UKQCD spectrum
results obtained using this action. I compare his $\rho$~mass with masses
computed by Weingarten and collaborators\,\cite{donW}
using the Wilson quark action in
Figure~\ref{rho-sw}. I use the QCD scale parameter $\Lambda_V$ determined
from the measured value of $\alpha_V(3.4/a)$ as a reference
scale.\footnote{Some
people prefer to use the string tension or other infrared quantities to fix
relative scales. The problem with that approach is that
most such quantities have
significant $a^2$~errors themselves, unless they are highly corrected as
are the NRQCD charmonium and upsilon spectra. $\Lambda_V$, on the other
hand, when {\em defined} in terms of the plaquette has no $a^n$~errors. The
only errors in $\Lambda_V$ are from higher-order perturbation theory, and
these are negligible.} Results from the Wilson quark action
show large finite-$a$ errors, while the SW~results are $a$-independent to
within (large) statistical errors. The two calculations agree
within errors on the $a\!=\!0$~mass.
Note that tadpole improvement almost doubles the
coefficient of $\sigma\cdot F$ in the SW~action and so is very important to
the success of this action\,\cite{borici}.

These results show that the SW~quark action is probably sufficiently
accurate for most of our needs provided
$a$~is less than about~$1/6$~fm ($\beta\!=\!5.7$ or higher for quenched
simulations). However the
$a^2$~errors in this action become noticable at larger lattice spacings.
This is clear in the SCRI~results presented at this conference by Robert
Edwards for the
$\rho$~mass compared with the string tension. His results show finite-$a$
errors of order 20--25\% at $a\!=\!.4$~fm, which is consistent with my
estimate inspired by NRQCD analyses. Alford, Klassen and I noticed that the
leading correction to SW,
\be
\delta S = \frac{a^2}{6}\sum_\mu D^3_\mu\gamma_\mu,
\ee
violates relativistic invariance. We test for this by computing the square
of the speed of light using the pion's dispersion relation:
\be
c^2 = \frac{E_\pi^2(\pv) - m_\pi^2}{\pv^2}.
\ee
We find that $c^2$ is 30--40\% too small when we use the SW~action with
$a\!=\!.4$~fm\,\cite{d234}.

The SW~action must be replaced by a more accurate action before useful
simulations involving light quarks are possible at $a\!=\!.4$~fm. Alford,
Klassen and I are examining a variety of such actions. One of the
first we studied is the ``D234 action:''
\bearray
S_{\rm D234} &=& \sum \overline\psi\left\{-\Delta^{(1)}\cdot\gamma + m
\right.\nl  && -\frac{ra}{2}\left(\Delta^{(2)} + \frac{g\sigma\cdot
F}{2}\right) \nl
&& \left.+ \frac{a^2}{6} \Delta^{(3)}\cdot\gamma +
\frac{a^3}{12}\,\Delta^{(4)}
\right\}\psi,
\eearray
where $\Delta^{(3)}_\mu$ is a discretization of $D_\mu^3$, and
$\Delta^{(4)}$ is a discretization of $\sum_\mu D_\mu^4$. This action
made the error in $c^2$ three or four times smaller, and had several other
successes as well\,\cite{d234}. I worry that quark doublers introduced by
the higher-order differences may be a problem for this action, but there are
several ways of dealing with this problem that we are exploring.

I have no doubt that actions substantially
more accurate than the SW~action will be
available shortly. The actions discussed above are all of the ``perfect
field'' or Symanzik variety. Work has also begun on ``perfect actions'' for
the quarks, but is still in its infancy\,\cite{wiese}.

\subsection{Computing Cost}
The use of improved actions on coarse lattices should lead to substantial
reductions in simulation costs. My own experience is that the reductions are
spectacular. Almost all of the $a\!=\!.4$~fm results for gluons and heavy
quarks that I discuss in this paper were generated on an IBM~RS6000/250
workstation, which performs like a mid to high-range PC.
These simulations
typically involved lattices with only 6~points on a side, which fit easily in
the 8--12~MB of memory common on PC's. Quantities like the charmonium
wavefunction can be computed from scratch in an hour or two on such
a workstation;
all previous work on charmonium was done with nothing smaller than a CRAY,
and turnaround times were generally quite a bit longer than an hour.

Beyond the savings associated with smaller grids, a large part of the cost
reduction in these simulations is due to the near absence of critical slowing
down on coarse grids. Field configurations decorrelate after 10--20~sweeps
even with slow algorithms like Metropolis. Note also that the asymptotic
exponential behavior in propagators usually sets in after only one or two
lattice spacings; this is partly because operators on a coarse lattice
are automatically smeared (there are no high momentum states), and
partly because the asymptotic region occurs at distances of order
$1/2$--1~fm, which is only one or two lattice spacings on a coarse lattice.
The added cost of simulating with a more complicated action is negligible
(factor of two or three) compared with these savings.

\section{Conclusions}
The results I have reviewed here argue persuasively, I think, that
classically improved  actions can be made accurate to a
few percent for lattice spacings as large as~.4~fm.
Furthermore classical (i.e., tree-level in perturbation theory)
improvement is sufficient for this purpose, provided lattice operators
are tadpole-improved. Without tadpole improvement, finite-$a$~errors are only
half removed.

Some of the features of the gluon actions
I have discussed were anticipated
in actions obtained using numerical techniques like the Monte
Carlo Renormalization Group (see work by Gupta, Iwasaki and
others). But the high cost of these techniques greatly limits their
general utility. Renormalized perturbation theory and tadpole
improvement allow us to realize the same goals but with negligible costs.

Both gluon and heavy-quark actions
have been used with great success at~.4~fm, and work is proceeding on the
development of new, more accurate actions for the light quarks. Even without
new light-quark actions, there is now evidence that the
Sheikholeslami-Wohlert quark action is sufficiently accurate for
state-of-the-art work at $a\!=\!1/6$~fm ($\beta\!=\!5.7$ quenched QCD), which
is already a major advance.

As we shift to coarse lattices, numerical QCD will be redefined by a
$10^3$--$10^6$ speedup in simulations. This will have far reaching implications
for both the sociology of our field and the physics it produces. The
entry-level
computer needed for serious QCD simulations is no longer a CRAY; a personal
computer or desktop computer will do\,---\,lattice QCD is now accessible to
almost anyone. The cycle time for ideas is reduced from months to
hours\,---\,we
can dare to be much more adventuresome than we have in the past. In the short
term we should be capable of an abundance of 10\% accurate results for spectra,
form factors, decay constants, and all sorts of other phenomenologically
relevant
quantities. By using our new techniques in simulations on our largest
computers, we should finally obtain reliable callibrations of the various
approximations we make in lattice QCD (especially quenching). By
combining rapid simulations on coarse grids with analytic methods, like
strong coupling expansions, we may move closer to analytic solutions of
low-energy QCD\,---\,that is, nonperturbative QCD without computers.

\end{document}